\documentclass[]{interact}
\usepackage{longtable}
\usepackage{ltablex}
\usepackage{lscape}
\usepackage{lipsum} 
\usepackage{graphicx}
\usepackage{relsize} 
\usepackage{makecell}
\usepackage{afterpage}
\usepackage{soul, color, xcolor}

\usepackage{rotating}

\usepackage{xltabular}
\usepackage{tabularray}
\usepackage{xcolor}

\usepackage{url}

\usepackage{gensymb}

\usepackage{enumitem}
\usepackage{epstopdf}
\usepackage[caption=false]{subfig}
\usepackage[numbers,sort&compress]{natbib}
\bibpunct[, ]{[}{]}{,}{n}{,}{,}
\renewcommand\bibfont{\fontsize{10}{12}\selectfont}
\makeatletter
\def\NAT@def@citea{\def\@citea{\NAT@separator}}
\makeatother

\theoremstyle{plain}

\theoremstyle{definition}

\theoremstyle{remark}

\sethlcolor{white}

\begin{document}

\title{A Review on Organ Deformation Modeling Approaches \hl{for Reliable Surgical Navigation} using Augmented Reality}

\author{
\name{Zheng Han and Qi Dou\thanks{Corresponding author: Qi Dou (qidou@cuhk.edu.hk).}}
\affil{Department of Computer Science and Engineering, The Chinese University of Hong Kong, Hong Kong, China}
}

\maketitle

\begin{abstract}

Augmented Reality (AR) \hl{holds the potential} to revolutionize surgical procedures by allowing surgeons to visualize critical structures within the patient's body. This is achieved through superimposing preoperative organ models onto the actual anatomy. Challenges arise from dynamic deformations of organs during surgery, making preoperative models inadequate for faithfully representing intraoperative anatomy. To enable reliable navigation in augmented surgery, modeling of intraoperative deformation to obtain an accurate alignment of the preoperative organ model with the intraoperative anatomy is indispensable. Despite the existence of various methods proposed to model intraoperative organ deformation, there are still few literature reviews that systematically categorize and summarize these approaches. This review aims to fill this gap by providing a comprehensive and technical-oriented overview of modeling methods for intraoperative organ deformation in augmented reality in surgery. Through a systematic search and screening process, 112 closely relevant papers were included in this review. By presenting the current status of organ deformation modeling methods and their clinical applications, this review seeks to enhance the understanding of organ deformation modeling in AR-guided surgery, and discuss the potential topics for future advancements.

\end{abstract}


\section{Introduction}

\begin{figure}[!htb]
\centering
\includegraphics[width=1.0\textwidth]{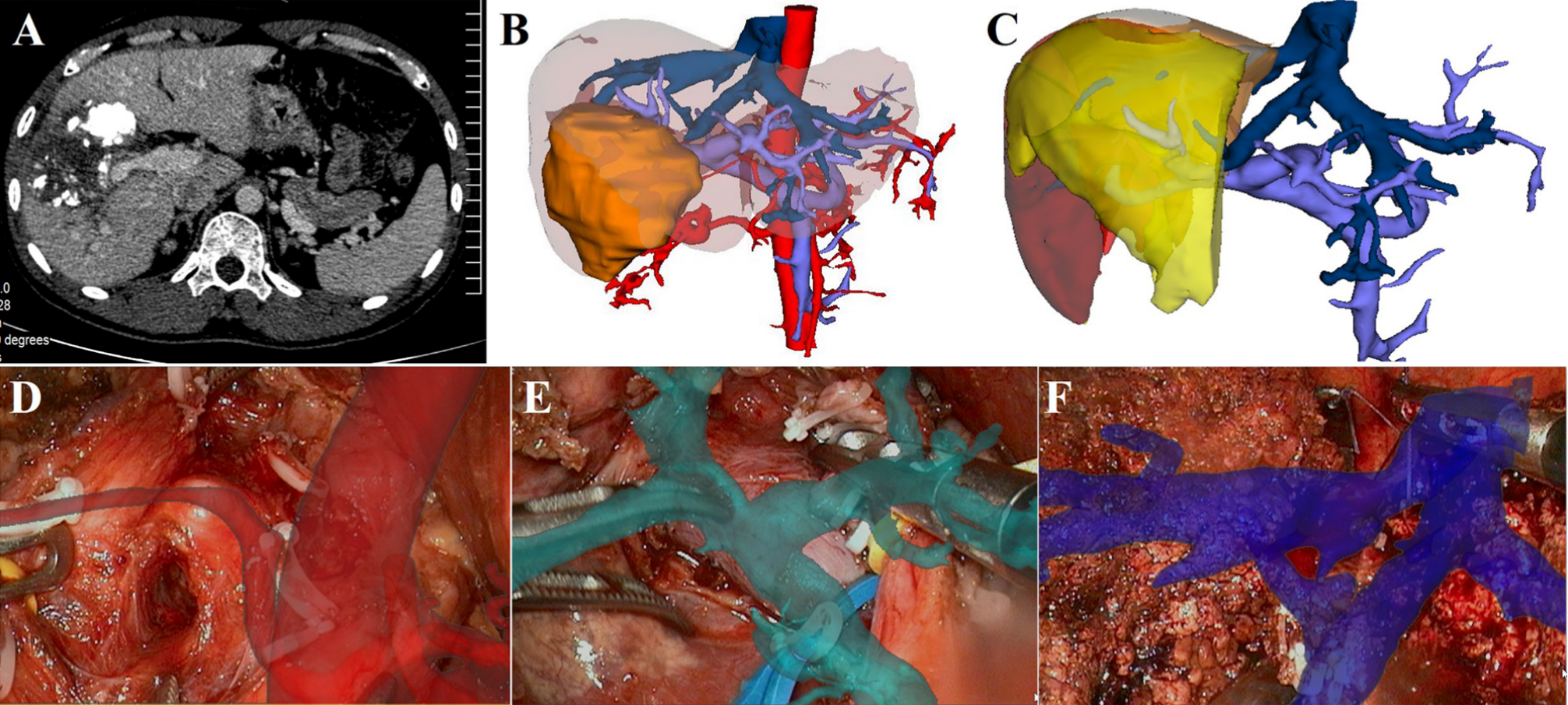}
\caption{Surgical navigation using augmented reality with an illustrative example of hepatectomy~\cite{zhang2021augmented}: A: preoperative CT images containing the liver; B: 3D reconstructed organ models for liver; C: residual liver volume from simulated right hepatectomy; D: intraoperative navigation of the hepatic artery; E: intraoperative navigation of the portal vein; F:  intraoperative navigation of the hepatic vein. Permissions: images licensed from the reference~\cite{zhang2021augmented} under CC BY 4.0.}
\label{AR_surgery}
\end{figure}

Recent advancements in optical see-through displays have introduced augmented reality (AR) as a promising tool for surgical navigation \cite{bernhardt2017status}. In AR-guided surgery, 3D digital organ models can be overlaid onto real anatomical structures \cite{birlo2022utility}. These organ models are generated from patient-specific data obtained through preoperative computed tomography (CT) or magnetic resonance imaging (MRI) images. They encompass information that presents organ shapes, surfaces, blood vessels, and tumors. The alignment of these digital models with patient anatomy helps surgeons to form an intuitive understanding of the spatial relationships among various anatomical structures concealed beneath the organ surfaces (see Figure \ref{AR_surgery}). \hl{The improved spatial awareness alleviates the burden for surgeons in localizing the safety-critical structures. This facilitates the execution of surgical plans with a higher level of precision and more informed decisions that could contribute to improved surgical outcomes}~\cite{luo2020augmented,chan2020augmented}.

Despite the potentials of AR-guided surgery, its reliability can be compromised by organ deformations caused by factors such as patient positioning, respiratory motion, extrinsic compression by pneumothorax, hematoma, or the device \cite{bernhardt2016automatic,abi2013image}. These deformations lead to misalignment between the preoperative digital organ models and the intraoperative anatomy, affecting the accuracy of localizing tumors and vessels \cite{chen2020tissue}. \hl{To make reliable surgical navigation, organ deformation modeling technique is an indispensable component in the AR-guided surgery system.} This modeling process involves continuously adjusting preoperative 3D organ models to adapt to the dynamic deformations occurring during surgical procedures \cite{furushiro2002specification}. Organ deformations can be observed through various modalities, including the locations of anatomical landmarks~\cite{chen2020tissue,shi2022synergistic}, tissue structure silhouettes \cite{peterlik2018fast}, or 3D digitized organ surfaces \cite{zhang2020assessment,li2019augmented}. Organ deformation modeling, at its core, involves extrapolating organ deformation fields based on these intraoperative observations to adjust preoperative digital organ models.

While organ deformation modeling plays a pivotal role in AR-guided surgery, achieving consistency between digital models and real organs remains a persistent challenge due to technical constraints. Firstly, organ surfaces often lack distinct features suitable for use as fiducials in deformation modeling, and the limited color and texture contrast in intraoperative imaging modalities further exacerbate this issue \cite{lathrop2009conoscopic,dos2014pose}. Secondly, the intraoperative observations available for deformation modeling are relatively limited, typically allowing only partial views of organs to be acquired \cite{reichard2017projective}. Extrapolating deformations spread across the entire organ from such limited information poses a significant challenge, especially given the considerable deformation that soft tissues within organs may experience during surgical procedures \cite{jia2021improving}. Thirdly, different surgical specialties present unique challenges. For instance, organs such as livers and kidneys exhibit viscoelastic deformation behavior during surgery \cite{nava2008vivo,boubaker2009finite,tillier2006finite}, while spinal structures deform due to their inherent flexibility \cite{kadoury2011automatic}. In these regards, specialized deformation modeling methods should be tailored for each surgical specialty to effectively address the corresponding organ deformations.

To inspire potential solutions for addressing these challenges in the computer-assisted surgery research field, this literature review commences by providing a comprehensive and technical-oriented summary of current deformation modeling techniques (cf. Sec. 3). Subsequently, it delves into various surgical specialties, offering detailed insights into the particular challenges faced by each specialty and presenting existing solutions (cf. Sec. 4). Finally, drawing on existing research, this review discusses future research possibilities (cf. Sec. 5).

\section{Literature Search and Screening Process}

\begin{figure}[!htb]
\centering
\includegraphics[width=1.0\textwidth]{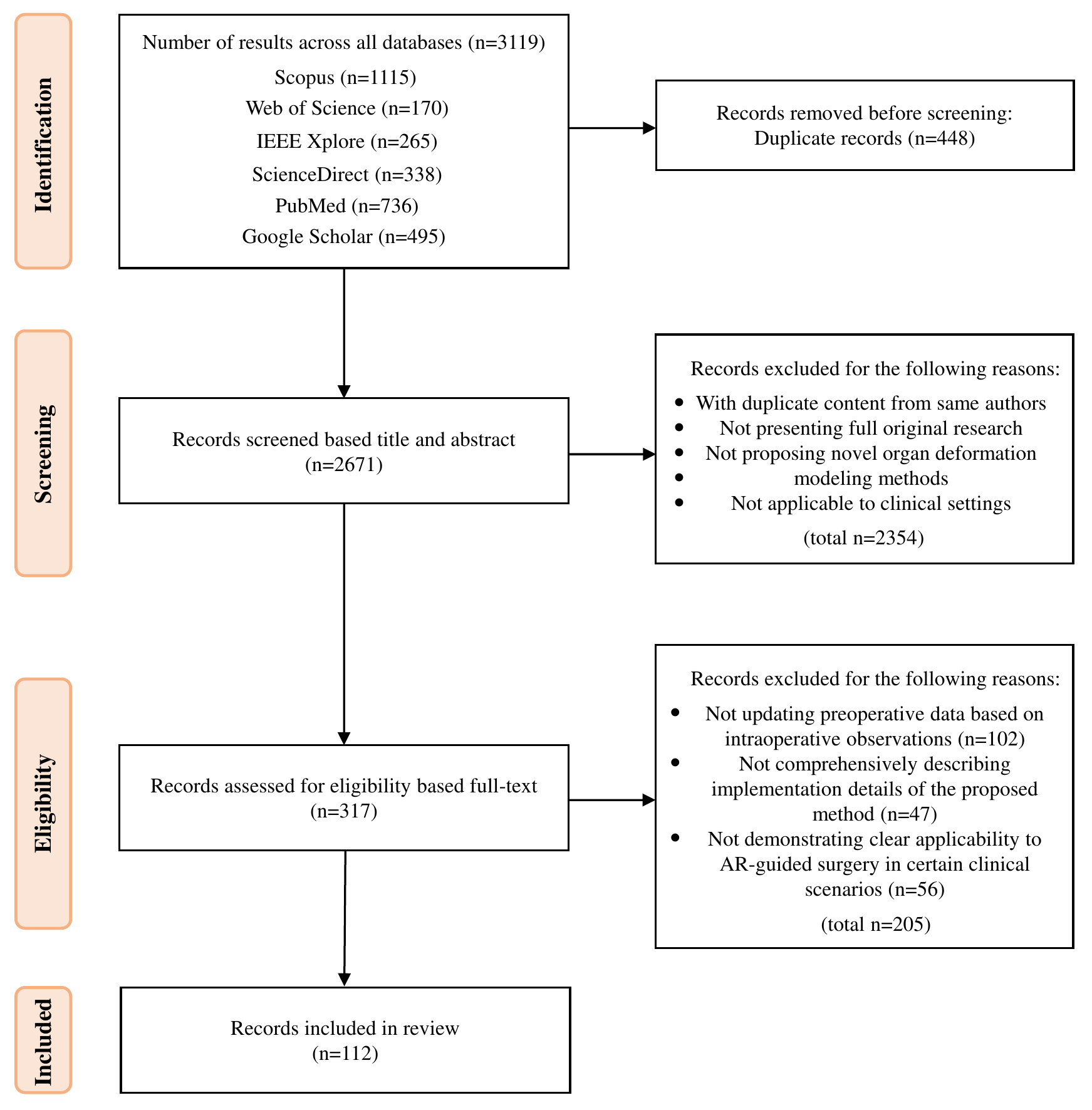}
\caption{The workflow of the screening process using PRISMA~\cite{page2021prisma} protocol for the performed review.}
\label{PRISMA_workflow}
\end{figure}

The literature search and screening process were conducted according to the guideline of \emph{Preferred reporting items for systematic review and meta-analysis (PRISMA)}~\cite{page2021prisma}. Figure~\ref{PRISMA_workflow} shows the overview of our literature processing workflow.

\subsection{Literature search}
The literature search was initially conducted in May 2023, across five scientific databases: Scopus, Web of Science, IEEE Xplore, ScienceDirect, and PubMed. \hl{An updated search on Google Scholar for 2023 and 2024 articles was subsequently performed on in March 2024 in order to add more literature including ArXiv papers.} To make a comprehensive investigation, a number of search terms was identified, comprising keywords including ``intraoperative", ``non-rigid", ``deformation", ``registration", ``modeling", and ``AR". Logical operators (AND/OR) were utilized to combine the keywords, facilitating a thorough and proper retrieval of relevant articles from each database. A detailed list of these search terms can be found in Appendix Table \ref{tbl:table_appendix}.

\subsection{Selection process}
The literature search yielded a total of 3119 records, and after removing duplicates, 2671 unique studies underwent screening. During the screening phase, titles and abstracts were reviewed to preliminarily exclude irrelevant publications. The exclusion criteria comprised the following items: (1) studies containing similar content from the same authors, (2) non-original research such as reviews or book chapters, (3) studies not focusing on modeling organ deformation, (4) studies not proposing new methods (e.g., comparative studies evaluating existing methods), and (5) studies not validated in scenarios related to clinical settings (e.g., focused solely on medical simulations or virtual surgical scenarios). Based on these criteria, 2354 records were excluded.

Subsequently, a more time-consuming full-text assessment was conducted to determine the eligibility of the remaining 317 studies for inclusion to this survey. Full-text articles had to satisfy the following inclusion criteria: (1) updating preoperative reconstructed 3D organ models based on information extracted from an intraoperative acquisition, (2) providing a comprehensive description of the proposed method's implementation details, and (3) demonstrating clear applicability to AR-guided surgery in certain clinical scenarios. After applying these criteria, a final set of 112 studies were included in this review paper. In this way, we hope that the screened compact set of references can present the most closely-related literature and highly-informative summary, so that readers can find a focused survey on the specific topic of organ deformation modeling approaches in AI-guided surgery.

\subsection{\hl{Related review papers}}
To ensure new knowledge provided by this review, we also examine existing survey papers on related topics identified during the literature search and screening process.

\citet{min20233d} (2023) review registration approaches for aligning preoperative organ models with intraoperative anatomy. However, their focus is solely on orthopedic surgery and thus does not cover the deformation of soft tissues. Similarly, \citet{gsaxner2023hololens} (2023) summarize methods for rigidly aligning digital and physical coordinates to realize surgical navigation in augmented reality space, but compensation for soft tissue deformation is still lacking. \citet{schneider2021performance} (2021) provide a comprehensive comparison of current surgical navigation systems with augmented reality support, regarding accuracy performance in laparoscopic liver surgery. While they discuss the issue of soft tissue deformation, a technical overview of how to model organ deformation is not provided. The most pertinent review on organ deformation modeling techniques is by \citet{bernhardt2017status} (2017), who summarize non-rigid registration methods to account for soft tissue deformation. However, this review was published a while ago, therefore a newer survey is needed to update the latest literature especially learning-based and data-driven techniques. 

\section{Organ Deformation Modeling Approaches}
\begin{figure}[!htb]
\centering
\includegraphics[width=1.0\textwidth]{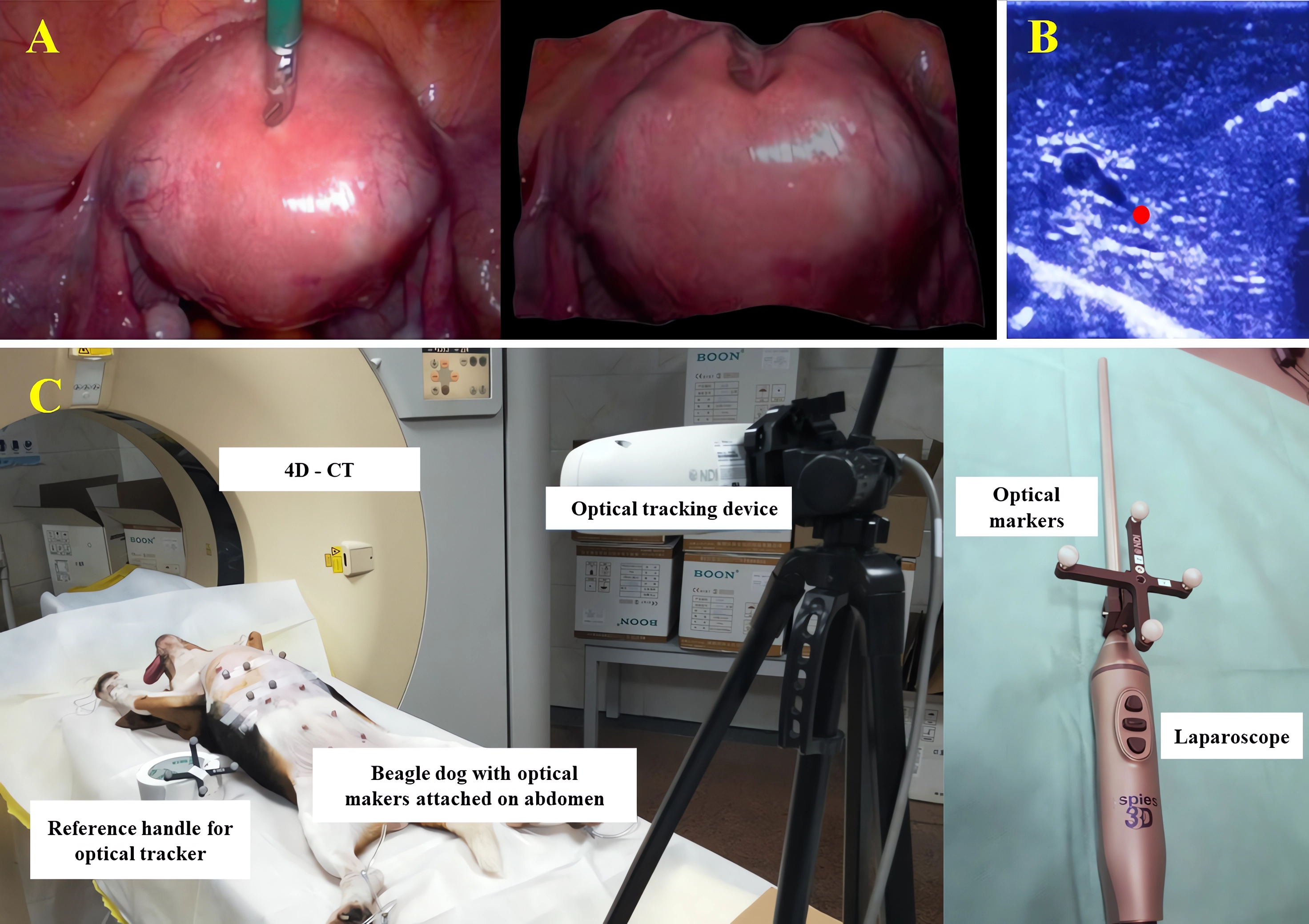}
\caption{Illustration of several common intraoperative observations. A: reconstructed organ surface (right) from laparoscopic images (left); B: the position of anatomical points (red points) manually picked by surgeons on Ultrasound images; C: optical markers attach on the abdomen (left) and instruments (right). Permissions: some illustrative images reprinted from representative references~\cite{maier2013optical,maier2013optical,bernhardt2017status} under respective copyright license permission from Elsevier.}
\label{Observations}
\end{figure}

Organ deformation modeling involves predicting a deformation field as output to update a pre-surgery reconstructed 3D organ model based on the input intraoperative observations~\cite{furushiro2002specification}. These observations can be anatomical point coordinates, or partial organ surface reconstructed from stereoscopic images, as illustrated in \hl{Figure} \ref{Observations}. By incorporating these observations, the modeling process ensures the fidelity of preoperative organ models to intraoperative anatomical changes, thus accommodating tissue deformations during AR-guided surgery. \hl{With the potential to enhance the surgical precision, this field has attracted widespread attention, leading to the flourishing of various modeling algorithms.}

\subsection{Categorization of organ deformation modeling methods}
Existing algorithms for modeling organ deformations can be broadly categorized into model-based, data-driven, and hybrid methods. Table \ref{tb:Model-based deformation modeling methods} and Table~\ref{tb:Data-driven and hybrid methods} provide a categorical summary of these existing methods for organ deformation modeling.

\subsubsection{Model-based methods}
In the field of organ deformation modeling, the majority of research efforts were initially directed towards model-based algorithms. These early research endeavors centered around understanding the tracking and modeling of tumor movement within organs during respiratory motion patterns. Clinical trials conducted by \citet{schweikard2000robotic} confirmed the hypothesis that a correlation exists between internal and external motion. This discovery laid the foundation for the development of correlation models, which could predict the intraoperative motion of internal tumors based on changes in externally trackable signals, often referred to as surrogates. Correlation models found practical application in cases such as radiofrequency ablation therapy \cite{li2019augmented}, where small tumors could be treated as rigid targets. However, when dealing with the deformation of soft tissues, correlation modeling proved to have limitations in its fitting capacity, presenting challenges \cite{mcclelland2013respiratory}.

To achieve the goal of modeling soft tissue deformations, further exploration ensued. Existing model-based algorithms can be categorized into four subcategories based on their underlying principles: (1) biomechanical models, (2) physics-based modeling methods, (3) geometry-based alignment methods, and (4) statistical models.

Biomechanical models (\textit{n = 44}) account for soft tissue deformations by numerically solving partial differential equations associated with constitutive models \cite{mendizabal2020physics}. Finite element (FE) methods are commonly used in the numerical solving process due to their effectiveness in handling partial differential equations (PDEs). In biomechanical modeling, the problem of inferring organ deformation is framed as solving a boundary value problem, where the boundary conditions \hl{(input data)} are derived from intraoperative observations \cite{reichard2017projective}. Boundary conditions (BCs) on an FE model can be expressed as either displacements or forces. When specifying a displacement BC, a node is compelled to move to a given position. Conversely, when a force BC is specified, it ensures that the internal stress is in equilibrium with the applied stress. With specified BCs, biomechanical models have demonstrated promising capabilities in simulating the viscoelastic behavior of soft tissues within various organs, such as the liver, prostate, kidney, and brain \cite{nava2008vivo, boubaker2009finite, tillier2006finite, kemper2004anisotropic}. However, their application in a real surgical setting encounters difficulties. \hl{Firstly, acquiring BCs during surgery remains challenging. Obtaining displacement BCs requires known surface correspondences, while obtaining force BCs necessitates precise measurement of forces applied by surgical tools on the organ} \cite{pfeiffer2020non}. \hl{Secondly, directly solving FE-based biomechanical models presents logistical limitations, such as the lengthy computational time} \cite{clements2007atlas}. These challenges limit the practical utility of biomechanical models in clinical scenarios. To address the issues, \citet{yang2024boundary} recently proposed a \hl{boundary constraint-free biomechanical model}, eliminating the need for predefined zero-displacements and force locations in BCs. \citet{min2023non} explored the use of \hl{physics-informed neural networks} to solve PDEs, offering potential solutions to computational efficiency challenges.

Physics-based modeling (\textit{n = 6}) presents an alternative approach to soft tissue deformation modeling. The fundamental concept behind these methods is to depict organ deformation in a physically interpretable manner. For instance, in the work by \citet{suwelack2014physics}, the non-rigid deformation problem is likened to an electrostatic-elastic scenario, where an elastic representation of the preoperative liver model behaves akin to an electrically charged object interacting with the oppositely charged rigid intraoperative liver surface. The preoperative liver in its undeformed state and the intraoperative liver surface, captured through stereo endoscopic imaging, \hl{serve as inputs} for the registration process. Electrostatic forces facilitated the alignment of the preoperative liver with the intraoperative surface, with elastic forces providing regularization. Another approach, demonstrated by \citet{dagon2008framework}, involved modeling hepatic vein deformations using a mass-spring skeleton aligned with vessel centerlines. During surgery, the actual vessel positions were inferred from image segmentation provided by a tracked 2D ultrasound system. These intraoperative measurements were then translated into 3D points, serving as input data. Elastic deformation of the skeleton to align with these 3D points was achieved by applying virtual forces until it conformed to the measurements. Currently, research in physics-based modeling remains relatively limited, and there is a need to explore additional physical constraints to enhance the realism of deformation modeling \cite{costa2012novel}.

Geometry-based alignment methods (\textit{n = 21}) handle 3D organ models in the form of point clouds. These methods deform the preoperative 3D organ model by manipulating its geometric coordinates, specifically the 3D points that compose the organ surface \cite{schneider2020comparison}. They aim to find a rigid transformation while simultaneously deforming the preoperative 3D model \hl{(output)} to align it with the intraoperative organ surface point clouds \hl{(input data)}. This is achieved by employing non-rigid point cloud registration algorithms, such as multi-stage iterative closest point (ICP), coherent point drift (CPD), or thin-plate spline (TPS) registration \cite{chen2021method, zhang2020assessment, maris2017deformable}. These algorithms estimate point correspondences and transformation functions between preoperative and intraoperative surfaces, aiming to maximize the alignment of visible regions between preoperative and intraoperative surfaces \cite{ma2022augmented}. However, rather than FE-based or physics-based approaches, geometry-based alignment methods often lack strain energy regularization \cite{fung1984structure} in the deformation modeling process. This regularization, described by the material properties and geometric features of organs, ensures that the deformation results adhere to organs' elastic properties and mechanical behavior. \hl{The absence of this regularization in geometry-based alignment methods can diminish the accuracy and reliability of the predicted deformations in invisible surface regions.} Moreover, geometry-based alignment methods face challenges in effectively propagating external organ surface deformation to internal structures such as vessels and tumors. In a preliminary study conducted by \citet{maris2017deformable}, a direct application of the non-rigid TPS function obtained from the surface registration to each of the points \textit{x} of the target tumor was tested, resulting in an evaluation error of around $5 mm$ on the centroid of the tumor volume.

Statistical models (\textit{n = 9}) describe the patterns of organ shape variations or motion from a statistical perspective, corresponding to statistical shape models and statistical motion models. In the shape model, variations among shapes from different individuals are accounted for, while the motion model captures the temporal changes in shape relative to a reference, such as those caused by respiratory motion \cite{jud2017statistical}. Constructing statistical models depends on having patient ground truth data, ideally offering high-resolution 3D spatial representations of organs. Particularly in the case of building statistical motion models, it is also crucial to have a sufficiently high temporal frame rate, capturing several volumes per second. To address the challenge of balancing image quality and acquisition speed, a learning-based technique was introduced by \citet{von20074d} to obtain 4D-MRI data. With the 4D-MRI data, a population-based statistical model can be built from the non-rigid registration of the MRI images \cite{rueckert1999nonrigid, modat2010fast}. The constructed statistical models can be employed to estimate present shape changes based on intraoperative observations. For instance, if online ultrasound images depict parts of the changing shape of interest, the full changing shape can be estimated form the previously observed shape changes, which have been encapsulated in the statistical motion model.

\hl{In general, model-based methods employ explicit mathematical models to characterize organ deformation. Among these, biomechanical models demonstrate good simulation effectiveness for deformations induced by external forces, such as instrument interactions and pneumoperitoneum pressure. For deformations arising from regular motion patterns, such as those caused by respiratory motion, statistical models are more suitable. Physics-based modeling endeavors to transfer existing physical models to directly simulate the viscoelastic behavior of soft tissues. Geometry-based alignment methods, on the other hand, tend to utilize non-rigid registration techniques from general computer vision to simultaneously address organ rigid registration and deformation modeling. However, this approach often lacks strain energy regularization, thereby failing to ensure that the predicted results adhere to organs' elastic properties and mechanical behavior.}

\begin{figure}[!htb]
\centering
\makebox[\textwidth][c]{\includegraphics[width=1.0\textwidth]{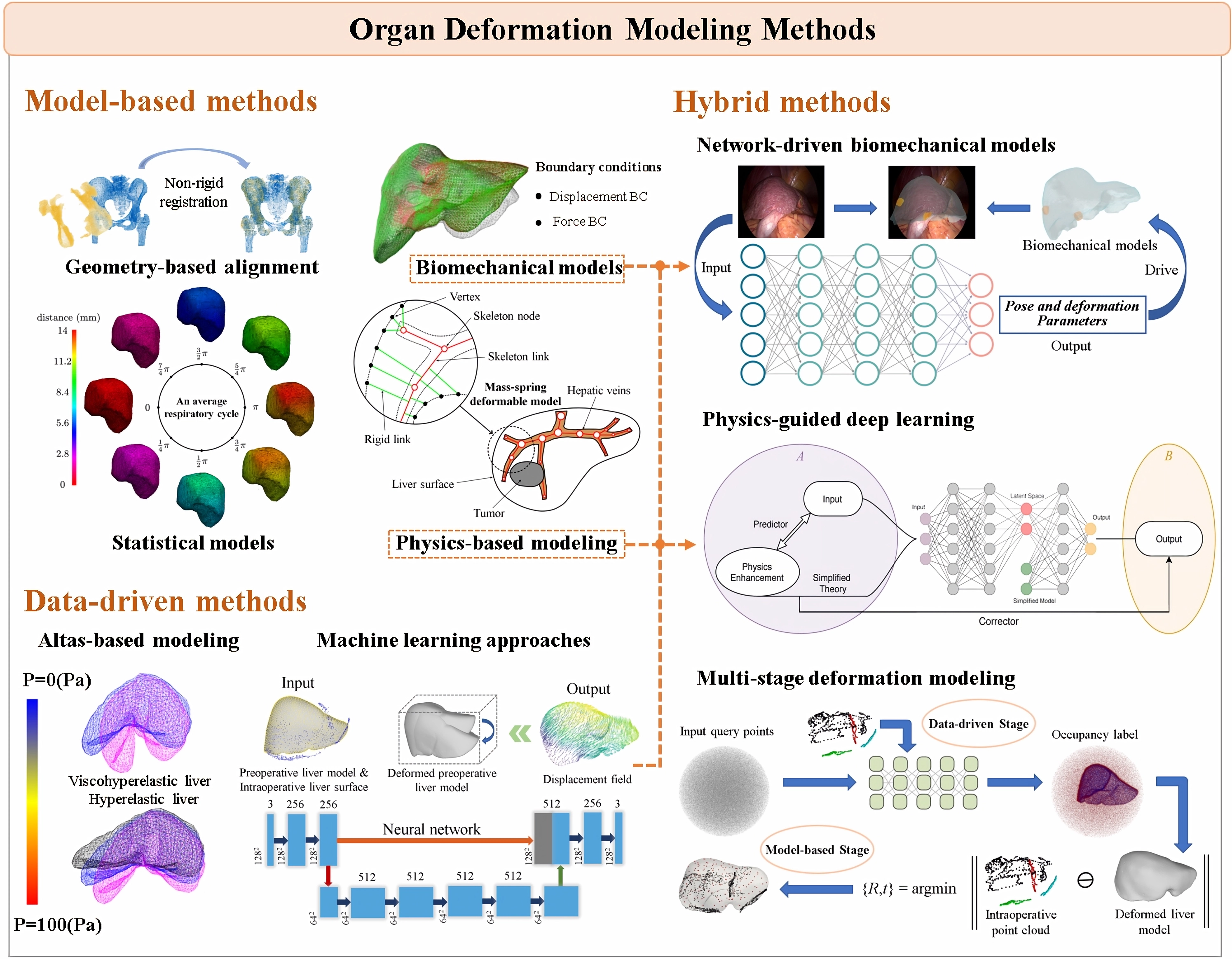}}
\caption{Overview of deformation modeling methods: (a) Model-based methods include: geometry-based alignment methods, biomechanical models, statistical models, and physics-based modeling. (b) Data-driven methods include: alras-based modeling, and machine learning approaches. (c) Hybrid methods include: network-driven biomechanical models, physics-guided deep learning, and multi-stage deformation modeling methods.
Permissions: some illustrative images reprinted from representative references~\cite{wang2023bone,mendizabal2020simulation,jud2017statistical,dagon2008framework,labrunie2023automatic,pawar2021physics,heiselman2024image,marchesseau2017nonlinear} under respective copyright license permission from Elsevier, IEEE, CC BY 4.0, or AIP Publishing.}
\label{Overview}
\end{figure}

\subsubsection{Data-driven methods}
Data-driven methods were initially employed to address the limitations of directly applying biomechanical models in surgical procedures. Specifically, early data-driven methods sought to reduce intraoperative computation of biomechanical models by constructing a patient-specific atlas generated by FE-based simulation \cite{clements2007atlas}. Recently, there has been a growing trend in adopting machine learning approaches as implicit representations of the underlying biomechanical model mechanism. The advantage of this adoption lies in the ability of machine learning approaches to predict the complex behavior of elastic organ structures in real-time without relying on preassigned boundary conditions \cite{shi2022synergistic}. This is particularly beneficial for clinical applicability, as boundary conditions are often challenging to obtain in clinical settings. The following provides an overview of the utilization of (1) atlas-based modeling approaches and (2) machine learning approaches in deformation modeling.

The atlas-based method (\textit{n = 5}) entails the construction of a pre-operatively computed collection of solutions, referred to as an atlas. This atlas is then used to align the data acquired during surgery with the solutions within the atlas, enabling the prediction and correction of the intraoperative organ deformation \cite{chen2010intraoperative}. The process of atlas construction begins with the generation of a patient-specific finite element organ model. This model is derived from the surface description provided by a segmentation of the pre-operatively obtained image volumes. Boundary conditions, patient orientations (e.g., gravity directions), and material properties are selected based on a priori knowledge of surgical loading conditions \cite{chen2013integrating}. The preoperative FE organ model is then run for each combination of conditions to create the atlas of organ deformation solutions. During surgery, a set of weight parameters \hl{(output)} is determined to linearly combine the pre-operatively computed collection of solutions within the atlas \cite{heiselman2018characterization}. This allows the organ model generated from atlas to match with the intraoperative observations, such as 3D digitized surfaces of the organ \hl{(input)}. These weight parameters can be iteratively solved using various algorithms, such as the iterative closest atlas algorithm \cite{clements2007atlas} or the Levenberg–Marquardt nonlinear optimization method \cite{lourakis2005brief}.

Machine learning approaches (\textit{n = 10}) were initially harnessed to expedite biomechanical modeling \cite{phellan2021real}. These models learn a function that maps \hl{inputs}, such as external forces, to \hl{outputs}, such as nodal displacements, by training on organ deformation datasets \cite{niroomandi2013real}. Typically, these datasets consist of synthetic data generated from simulating the biomechanical behavior of organs using FE methods \cite{zhu2022real}. Training the network using synthetic datasets rather than real patient data is a trade-off. Collecting ground truth deformation patterns of the same patient's organs would require multiple CT scans, which is impractical in clinical practice. Notably, studies such as \cite{saeed2020prostate, mendizabal2020physics, brunet2019physics, pfeiffer2019learning, pfeiffer2020non} have shown that even when trained solely on synthetic data, machine learning models still have the  potential to accurately predict the physical deformation behaviors of organs. After training, the machine learning model serves as an implicit representation of the underlying biomechanical model mechanism, eliminating the need for explicit mathematical formulations during inference. Further, machine learning approaches can be categorized into two main branches: traditional machine learning (TML) and deep learning (DL) techniques. Specifically, traditional machine learning techniques, such as support vector machines and random forests, have demonstrated their potential to simulate tissue behavior in real-time, including organs such as breasts \cite{martinez2017finite} and livers \cite{pellicer2020real,lorente2017framework}. However, these methods face limitations when applied in actual surgical settings. Traditional machine learning models rely on FE-based conditions, specifically stress and displacement conditions \cite{carter2005application}, as \hl{input variables}. Acquiring these variables in a real surgical environment is challenging, since, for instance, measuring forces requires additional devices, and estimating surface displacements necessitates knowledge of surface correspondences \cite{pfeiffer2020non}. Currently, the most widely adopted solution for obtaining input variables involves simplified boundary conditions, which assume fixed subsets of nodes while leaving others free \cite{plantefeve2016patient,lister2011development,plantefeve2014atlas}. However, manually assigning these boundary conditions can lead to instability if incorrect values are chosen, thereby impeding prediction accuracy. Deep learning, on the other hand, has been recognized as a powerful approach for predicting organ deformation behaviors, offering enhanced clinical applicability due to its ability to operate in real-time and independence from preassigned boundary conditions. These algorithms function by taking preoperative organ models, along with intraoperative observations such as 3D reconstructed organ surfaces \hl{(input)}, and predicting a displacement field \hl{(output)} that warps the preoperative models to align with these intraoperative observations. For instance, \citet{nakao2022image} introduced a deep learning-based framework for modeling the deformation of abdominal soft organs. This framework offers an end-to-end solution for real-time 2D/3D deformable registration by integrating an image-based generative network \cite{ronneberger2015u} and a graph convolutional network (GCN) \cite{kipf2016semi}. The generative network learns the transformation from the 2D projection image to a displacement map, while the GCN translates this transformation into the final nodal displacements of the organ model. Moreover, \citet{pfeiffer2020non} harnessed a fully 3D convolutional architecture to recover the displacement field directly from the intraoperative digitized organ surface, which was reconstructed from the laparoscopic video. This estimated displacement field can subsequently be utilized to deduce the nodal displacements spanning the entire organ, encompassing organ surface, vessels, and tumors. \hl{Table} \ref{tb:Public available datasets} \hl{summarizes publicly available datasets that can be used for neural network training and quantitatively evaluating accuracy.}

In general, data-driven models, by learning a vast number of deformation patterns, can serve as implicit representations of the underlying biomechanical mechanism. Thanks to parallel computing, the inference process of data-driven methods can be significantly accelerated \cite{mendizabal2020simulation}. Additionally, both atlas-based methods and deep learning approaches can directly infer deformation fields from intraoperative observations, greatly enhancing clinical applicability \cite{pfeiffer2019learning, pfeiffer2020non, mendizabal2020physics}. In contrast, traditional machine learning methods still rely on manually assigning boundary conditions, which can lead to instability and affect accuracy if incorrect values are chosen.

\subsubsection{Hybrid methods}

Hybrid methods represent the fusion of model-based and data-driven approaches, offering notable flexibility in their implementation.

One approach to implementing hybrid methods involves leveraging network-driven biomechanical models (\textit{n = 4}). This approach utilizes neural networks to estimate essential prerequisites, such as boundary conditions or deformation parameters, to drive biomechanical organ models. These  prerequisites are typically challenging to directly measure during actual surgical procedures. An example to better illustrate this concept is the work of \citet{tagliabue2021intra}, who employed a BANet \cite{tagliabue2021data} to continuously estimate boundary conditions from raw intraoperative point cloud data of the deforming anatomy, captured by a vision sensor. The estimated boundary conditions from BANet guide preoperative biomechanical organ models to deform, ensuring an accurate representation of intraoperative organ deformations. Another example comes from the work of \citet{labrunie2023automatic}, where neural networks predict pose (\textit{R, T}) and deformation parameters ($\beta$) defining the intraoperative state of the organ. Initially, a ResNet-50 network \cite{he2016deep} extracts organ boundary features from mini-invasive images. These extracted features, along with the temporary pose (\textit{R, T}), and deformation parameters ($\beta$), serve as inputs to a regression network \cite{carreira2016human}. This network iteratively updates the pose (\textit{R, T}) and deformation parameters ($\beta$), which are subsequently employed to drive the preoperative biomechanical organ model for an accurate representation of organ deformations.

Another method for implementing hybrid approaches involves multi-stage deformation modeling (\textit{n = 10}), which breaks down the deformation modeling process into multiple stages and selecting appropriate methods for each. For example, \citet{shi2022synergistic} decomposed organ deformation modeling into organ surface and internal structure deformation modeling stages. Initially, an external-internal correlation model \cite{atluri1999analysis} is used to estimate organ surface deformation by tracking the displacement of external markers attached to the skin. Subsequently, neural networks propagate the surface deformation to internal structures, such as vessels and tumors. Meanwhile, \citet{jia2021improving} achieved deformation modeling by first recovering organ shape and then performing non-rigid registration. They utilized a point cloud occupancy network \cite{jia2020learning} to infer the complete organ shape from partial organ surfaces. Then, a correction algorithm utilizing the Levenberg–Marquardt nonlinear optimization method \cite{rucker2013mechanics} is applied for non-rigid registration based on the organ shape inferred from the network.

Implementing hybrid methods through physics-guided deep learning (\textit{n = 3}) presents another viable solution. This approach incorporates established physical principles, equations, and laws into the training and inference process of deep learning models to ensure that the models’ predictions align with physical phenomena \cite{wang2021physics}. Pioneering work in this field was introduced by \citet{min2023non}, who employed a novel deep learning approach integrating physics-informed neural networks (PINNs) with PointNet for non-rigid medical image registration. The key physical principle integrated into the deep learning model is a loss function term based on biomechanical constraints, which ensures that the estimated spatial transformation is biophysically plausible. This method represents prostate point displacements using PointNet\cite{qi2017pointnet}, while PINNs impose elastic constraints on the estimated displacements. Comparison experiments have illustrated that incorporating these physics-guided constraints significantly reduces target registration error, especially for patients with large deformations, thereby demonstrating superior performance and generalizability to new subjects. 

In general, hybrid methods effectively leverage the strengths of both data-driven and model-based approaches. They can improve data-driven predictions to better approximate the viscoelastic characteristics observed in actual organs, or expedite model-based methods in acquiring the necessary prerequisites for controlling organ models. The flexibility in implementing hybrid methods can also expand their applicability in real clinical settings. 

\subsection{\hl{Comparison of different deformation modeling methods}}
We would like to compare different deformation modeling methods based on their accuracy, computational efficiency, and ease of implementation, categorizing their performance as ``high", ``medium", or ``low" for each criterion. The ``accuracy" criterion evaluates the method's ability to accurately predict the deformation behavior of soft tissues; ``computational efficiency" indicates its suitability for real-time applications; and ``ease of implementation" reflects the method's practical applicability in clinical settings, considering factors such as the ease of model construction, and the accessibility of required intraoperative observations (input data) during clinical practice. 

\subsubsection{Accuracy comparative analysis}
In terms of accuracy, methods utilized biomechanical FE-models best explain the deformation behavior induced by external forces. This is primarily due to the fact that this method is grounded in continuum mechanics theory \cite{brunet2019physics}. Additionally, this method can leverage prior knowledge about the mechanical load applied to organs to reduce the complexity of the problem space. This prior knowledge can be integrated into constraints within the adjoint optimization scheme \cite{mestdagh2022optimal}, or decomposed into a set of localized point forces distributed over the active contact surfaces of the organ to control deformation responses \cite{ringel2023comparing}. Conversely, methods that solely rely on surface correspondence and lack mechanical constraints, such as geometry-based alignment methods in model-based approaches, and data-driven methods, often lack strain energy regularization within the objective function, leading to limited control over deformation field irregularities. Additionally, deep learning approaches in data-driven methods may exhibit voxelization artifacts, associated with data discretization procedures at convolutional layers \cite{heiselman2024image}. However, incorporating physics-guided loss functions for optimizing the parameters of the deep learning model has the potential to enhance accuracy in unseen deformation scenarios \cite{karami2023real}. For deformations arising from regular motion patterns, such as those caused by respiratory motion, statistical models perform better. This is because statistical models are directly constructed from patient ground truth data. These models estimate evolving shapes based on previously observed shape changes encapsulated within the statistical motion model.

\subsubsection{Computational efficiency comparative analysis}
Implementing parallel processing is key to ensuring computational efficiency. However, achieving this in FE-based methods poses challenges due to the memory access arrangement required for concurrent updates of matrix entries describing the physical state \cite{ljungkvist2015techniques}. To overcome this obstacle, machine learning approaches have been effectively leveraged, benefiting from the parallelization capabilities of modern GPUs, to accelerate FE models \cite{karami2023real}. By training on datasets generated from FE simulations, the machine learning model serves as an implicit representation of the underlying biomechanical model mechanism, eliminating the need for explicit mathematical formulations during inference. Similarly, deep learning models take advantage of parallelizing well on modern GPUs to extract complex patterns efficiently. As for other model-based methods, including statistical models, geometry-based alignment methods, and physics-based modeling, the algorithms themselves are not inherently complex and can be parallelized for computation. However, geometry-based alignment methods typically require an iterative optimization framework, which needs serial implementation, consequently diminishing some of their computational efficiency.

\subsubsection{Ease of implementation comparative analysis}
The ease of model construction and the clinical accessibility of input data will both impact the applicability of methods in real clinical settings. Ideally, algorithms that do not require additional equipment for acquiring input data will have higher clinical applicability. Altas-based modeling and deep learning methods in data-driven approaches, as well as physics-based modeling and geometry-based alignment methods in model-based approaches, can directly utilize organ surface information reconstructed from stereo 3D laparoscopy to predict organ deformations. Statistical models can also predict deformation behavior directly from 3D reconstructed organ surfaces; however, the construction of statistical models requires acquiring patient ground truth data at different respiration stages \cite{jud2017statistical}, thereby increasing the difficulty of the model construction phase. Moreover, biomechanical models require stress or displacement boundary conditions as input, which can be challenging to obtain in clinical practice. Acquiring stress boundary conditions necessitates measuring forces applied by surgical tools on the organ \cite{pfeiffer2020non}. Displacement boundary conditions can be obtained by measuring the displacement of interacting instruments \cite{mendizabal2020physics}, but this also requires additional equipment for spatial tracking of the instruments. Consequently, acquiring input data for biomechanical models is more demanding compared to other methods. Utilizing neural networks to estimate boundary conditions for driving biomechanical organ models holds promise for expanding their utility in clinical settings \cite{tagliabue2021intra}.

\begin{figure}[!htb]
\centering
\includegraphics[width=1.0\textwidth]{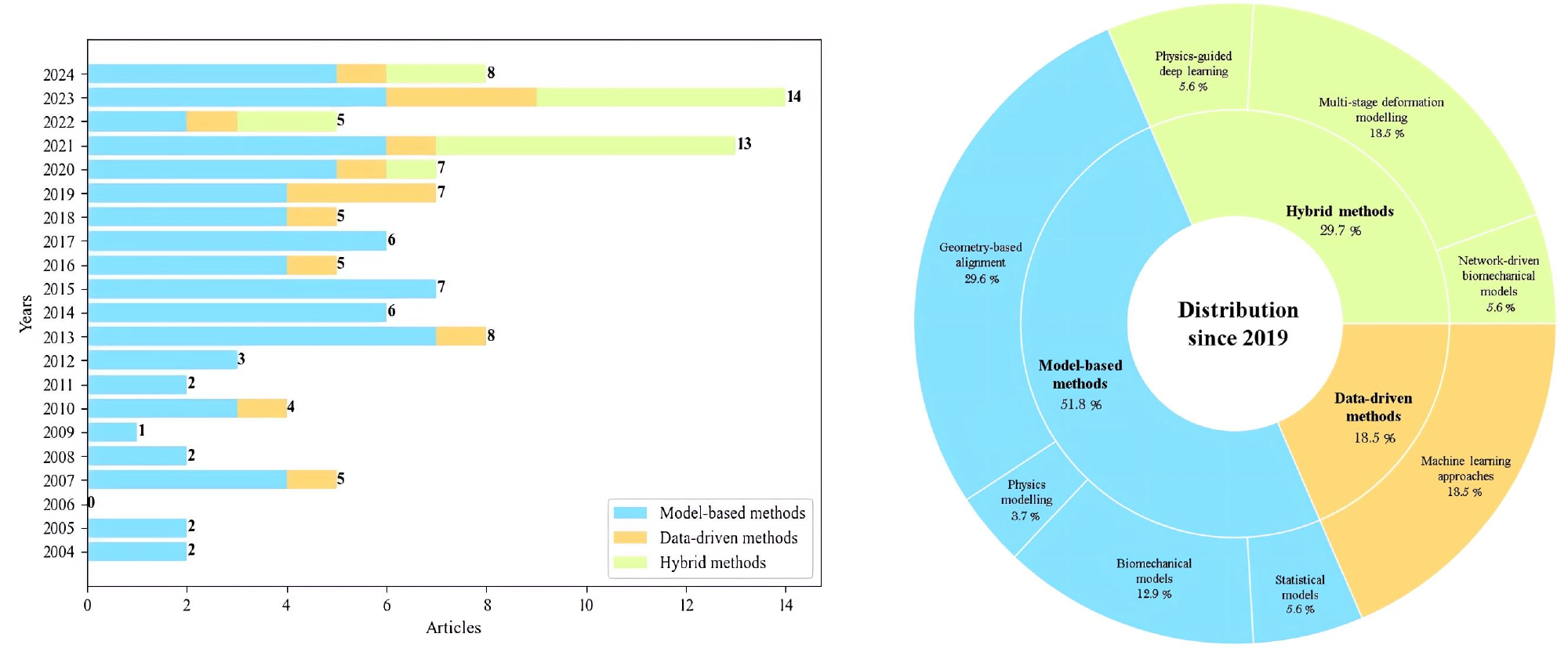}
\caption{Annual distribution of articles focusing on organ deformation modeling methods.}
\label{Annual distribution}
\end{figure}

\subsection{Analysis of annual distribution of relevant methods}

We further analyze the annual distribution of organ deformation modeling methods to understand their evolution over time. Figure \ref{Annual distribution} presents charts showing the annual distribution of articles focusing on these algorithms, revealing trends and patterns within the research landscape.

Between 2004 and 2018, research efforts primarily focused on model-based algorithms, underscoring their significance. Afterwards, starting in 2019, there was a noticeable increase in studies using neural networks to directly infer organ deformations from intraoperative observations, without manual parameter assignments. This shift led to the emergence of hybrid algorithms, combining both model-based and data-driven approaches. Notably, in 2020, there was a surge in articles on hybrid algorithms, reaching a comparable level with those on model-based algorithms. This rise reflects growing recognition of the benefits of integrating both approaches. Through a synergistic combination of model-based and data-driven methodologies, it is possible to refine data-driven results to better emulate the viscoelastic characteristics of real organs, or accelerate model-based methods in obtaining essential prerequisites for deforming organ models. As the field evolves, ongoing development of hybrid algorithms holds great promise for advancing the accuracy and efficacy of deformation modeling in clinical applications.

\section{Clinical Applications of Deformation Modeling for AR-guided Surgery}

\begin{figure}[!htb]
\centering
\includegraphics[width=1.0\textwidth]{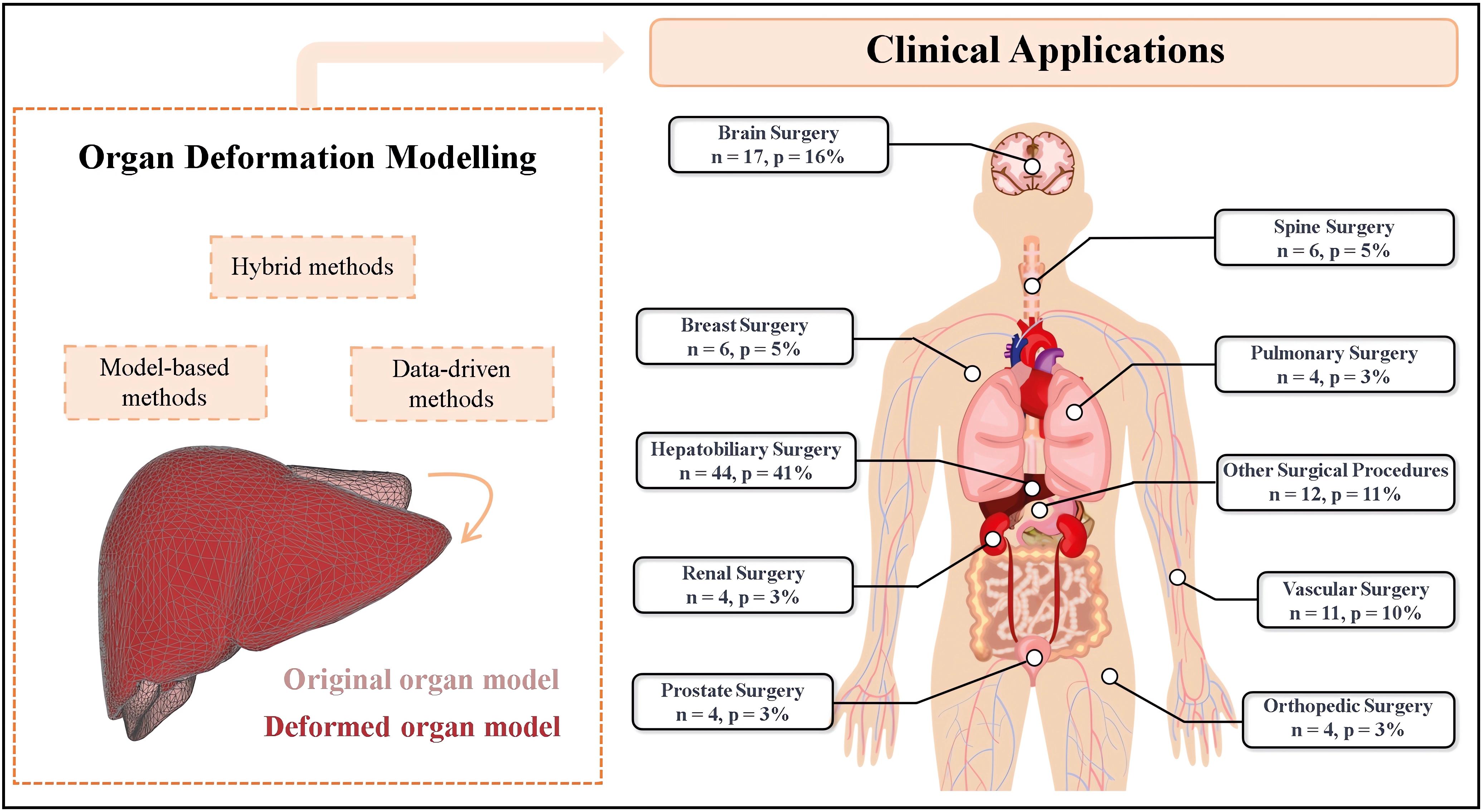}
\caption{Overview of reviewed 112 articles grouped by surgical specialty in clinical application scenarios.}
\label{Organ map}
\end{figure}

Organ deformation modeling has been explored across different surgical specialties to mitigate disparities between preoperative and intraoperative anatomical states, aiming to enhance the precision of surgical procedures. Figure \ref{Organ map} illustrates the distribution of research publications across these surgical specialties. Each specialty presents its own challenges, demanding tailored solutions and implementations for addressing them. The subsequent section will delve into the implementation of organ deformation modeling methods in each surgical specialty, including hepatobiliary surgery, brain surgery, breast surgery, spine surgery, vascular surgery, and renal surgery.

\subsection{Hepatobiliary surgery}

Organ deformation modeling finds applicability in hepatobiliary surgery procedures, including percutaneous tumor ablation and liver resection. 

In the context of percutaneous tumor ablation, the internal tumor position can be predicted in real-time from the external signals, such as the motion of trackable markers attached to the skin \cite{schweikard2004respiration}. This is achieved by establishing the correlation between the internal tumor position and external signals via regression or statistical tumor motion models \cite{li2021towards}. Furthermore, recent advancements \cite{shi2021internal, shi2022synergistic} have shown that even the deformation of healthy tissue near tumors, such as blood vessels, can also be predicted from external signals using neural networks. This empowers surgeons to effectively plan optimal puncture trajectories and perform tumor destruction while minimizing damage to adjacent healthy tissues. To date, in animal experiments applying liver deformation modeling, the exemplar achieved puncture accuracy is $3.52mm$ (pig) \cite{li2019mixed} and $2.50mm$ (dog) \cite{shi2022synergistic} in terms of targeting tumors, meeting the precision surgical safety requirement of an error less than $5mm$ \cite{nicolau2007clinical}.

\begin{figure}[!htb]
\centering
\includegraphics[width=1.0\textwidth]{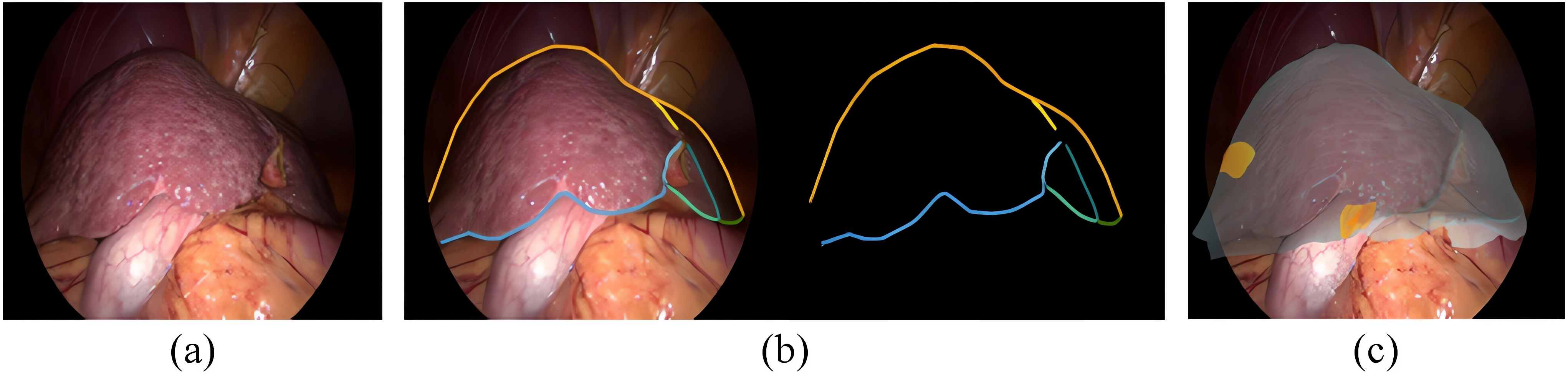}
\caption{Example of Liver deformation modeling pipeline \cite{labrunie2023automatic}. (a) Laparoscopic liver images; (b) extracted liver slihouettes; (c) projected deformed liver models. Permissions: images licensed under CC BY 4.0.}
\label{Liver silhouettes}
\end{figure}

In the context of liver resection, challenges arise when dealing with soft-tissue deformation modeling. Firstly, the traction exerted by surgical instruments induces large viscoelastic deformation in the liver organ, particularly in laparoscopic surgery, where pneumoperitoneum introduction further amplifies liver deformation \cite{luo2020augmented}. This deformation can greatly alter the spatial relationships among the liver's internal tissues \cite{bernhardt2016automatic}. Secondly, in liver resection procedures, available intraoperative information for liver deformation modeling is relatively limited. Typically, only a partial surface of the liver can be observed, often through laparoscopic cameras or specialized depth sensors designed for liver surface digitization \cite{golse2021augmented}. Inferring the deformation of the entire liver from this partial surface, especially given the lack of detailed texture information, presents a significant challenge \cite{lathrop2009conoscopic, pfeiffer2019learning}. To address these issues, \citet{pfeiffer2020non} attempted to predict the displacement field of the entire liver directly from the deformed partial liver surface, reconstructed from intraoperative laparoscopic video streams. The predicted displacement field can then drive the deformation of the entire preoperative liver model, including the liver surface, vessels, and tumors. Results from in silico experiments indicate a strong correlation between the accuracy of liver deformation modeling and the degree of liver deformation as well as the visibility ratio of the liver surface. \citet{labrunie2023automatic} pursued a similar approach to predict the liver's displacement field from laparoscopic video. However, instead of reconstructing the liver surface, they employed a ResNet-50 encoder \cite{he2016deep} to extract the liver's 2D slihouettes, thereby imposing further constraints on liver deformation modeling (as shown in Figure \ref{Liver silhouettes}). Accuracy experiments conducted in retrospective patient cases revealed a liver surface registration error of $8.5 mm$ in low deformation cases and $16.4 mm$ in cases involving large deformation.

\subsection{Brain surgery}

The primary approach for treating brain tumors is surgical resection, with the extent of resection being highly correlated to patient survival rates \cite{gonzalez2019state}. However, the phenomenon of brain shift, characterized by non-rigid tissue displacement attributed to factors such as cerebrospinal fluid drainage, tissue swelling, and gravitational effects, introduces misalignment between preoperative imaging and real-time intraoperative conditions. The misalignment may considerably affect the surgical outcome since neurosurgical procedures are often based on pre-operative planning where brain shift is not considered \cite{bayer2017intraoperative,miga2016clinical}.

\begin{figure}[!htb]
\centering
\includegraphics[width=1.0\textwidth]{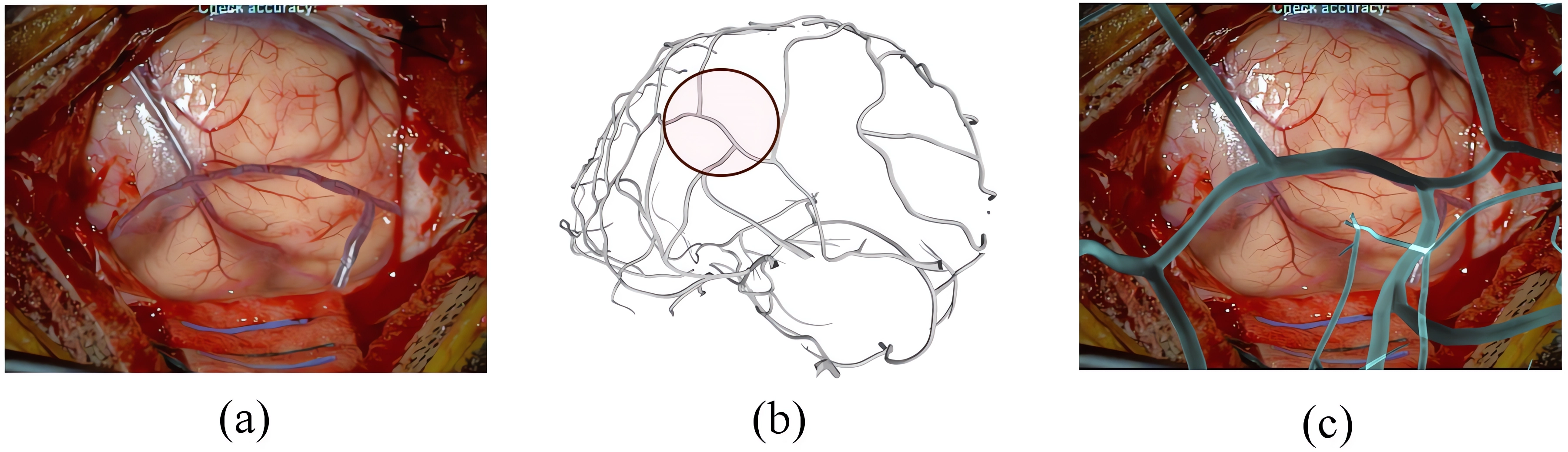}
\caption{Example of AR-guided brain surgery with cortical brain vessels \cite{haouchine2020alignment}. (a) intraoperative vessel image; (b) preoperative vessel model with region of interest; (c) pre- to intra-operative overlay with brain shift compensation. Permissions: images reprinted with permission from SPIE.}
\label{Brain vessels}
\end{figure}

Modeling brain deformation is essential to maintain alignment between preoperative surgical planning and intraoperative conditions. In craniotomy, this demands a high level of accuracy since any errors could potentially affect critical brain functional areas \cite{ghose2021automatic}. Notably, the contours of cortical vessels, exposed on the brain parenchyma surface, offer distinct features that are instrumental in achieving high-precise deformation modeling (as shown in Figure \ref{Brain vessels}). These vessel contours can be obtained through manual selection of points representing the starting and ending points of vessel segments \cite{haouchine2020alignment} or by directly employing neural networks to extract them from microscopic images \cite{haouchine2021pose, haouchine2020deformation}. Extracted cortical vessel features can serve as valuable inputs for driving biomechanical model deformations \cite{haouchine2020alignment} or as constraints for aligning preoperative models with intraoperative anatomy. In addition, some studies opt for an optimization approach to iteratively update the preoperative brain model, minimizing its displacement from the intraoperative 3D brain surface \cite{miga2016clinical,chen2013integrating,chen2010intraoperative}. The intraoperative 3D brain surface is typically segmented from intraoperative magnetic resonance images \cite{kemper2004anisotropic} or directly acquired using a laser range scanner \cite{zhuang2011sparse}.

Currently, the assessment of deformation modeling accuracy is primarily carried out under in-silico conditions. \citet{haouchine2021pose} reported impressive accuracy, achieving a target registration error below $1.93mm$ within the cortical level and immediate sub-cortical structures (depth $\leq$ $15mm$) during in-silico evaluations. The brain shift compensation demonstrated effectiveness of up to $68.2\%$, with a minimum compensation of $24.6\%$ even in the most challenging configurations \cite{haouchine2021pose}. However, errors exhibited an upward trend when targeting locations deeper within the brain (depth $\geq$ $30 mm$) or when confronted with greater degrees of deformation (depth $\geq$ $6 mm$) \cite{haouchine2021pose}.

\subsection{Breast surgery}

Diagnosis of breast cancer relies on mammography and preoperative MRI, during which the patient is positioned either standing or lying prone with pendant breasts. However, when performing breast conserving surgery, the patient is typically in a supine or semi-vertical position for ultrasound (US) guided interventions \cite{richey2022computational}. These shifts in patient positioning invariably result in breast deformation between the diagnostic and biopsy phases under the influence of gravity. Additionally, the compression forces from the US probe can cause significant deformation and repositioning of breast anatomy \cite{ferrari2023autonomous}. The cumulative effect of these factors presents challenges to surgeons in accurately correlating tumor locations between medical images and the surgical field.

To model breast deformation arising from alterations in patient positioning, registration can be driven using sparse data compatible with acquisition during breast conserving surgery \cite{richey2022computational}. This sparse data includes corresponding surface fiducials, sparse chest wall contours, and the intra-fiducial skin surface. For instance, \citet{ebrahimi2014using} observed deformations resulting from full abduction of the arm in supine MRI. They employed the thin plate splines registration algorithm, utilizing 24–34 fiducials evenly distributed across the breast surface, to rectify these deformations. Regarding deformation caused by the compression of the US probe, breast deformation can be predicted from the displacement of the US probe itself \cite{mendizabal2020physics}. Since the US probe is treated as a rigid body, it is reasonable to assume that when the anatomy is deformed by the probe during image acquisition, points on the breast surface beneath the probe will be displaced to the same extent as the probe. The displacement of surface nodes in contact with the US probe can then be used to predict displacements across the entire breast \cite{ronneberger2015u}. This method of relying on surface displacement inferred from the spatial tracking of the US probe, instead of directly monitoring surface deformations (e.g., via an RGBD camera), acknowledges that probe-induced deformations are significant but localized. Additionally, the probe itself typically obstructs much of the deformed surface from the sensor's view, rendering it challenging to directly measuring displacements of the contact surface \cite{mendizabal2020physics}.
A recent in-vivo experiment \cite{richey2022computational} involving seven different human subjects was conducted to assess the accuracy of breast deformation modeling. This evaluation, which focused on targeting an average of 22 subsurface markers, resulted in an average target registration error (TRE) of $6.3 mm$. 

\subsection{Spine surgery}

Spinal deformations occur during surgery due to the patient's respiration and surgical interventions, resulting in a disparity between preoperative CT scans of the spine and intraoperative conditions. Conventional intraoperative image-guided modalities, such as X-ray imaging, pose risks of ionizing radiation exposure for both patients and surgeons due to the need for multiple scans \cite{wang2019computer}. Addressing intraoperative deformations and radiation exposure are common challenges encountered in spinal surgery \cite{tanaka2021percutaneous}.

\begin{figure}[!htb]
\centering
\includegraphics[width=1.0\textwidth]{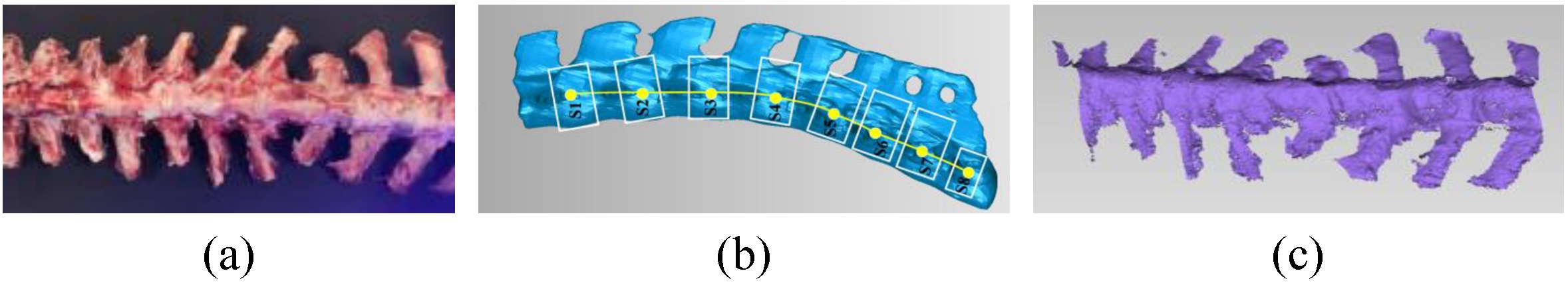}
\caption{Example of spine surgery with spine animal models \cite{chen2021method}. (a) The spine of a sheep; (b) preoperative 3D reconstruction and segmentation of spine based on CT; (c) intraoperative 3D reconstruction of spine based on binocular structured light. Permissions: images reprinted with permission from John Wiley and Sons.}
\label{Spine model}
\end{figure}

Modeling spine deformation without relying on conventional intraoperative imaging modalities presents potential as a solution. The approach to implementing spine deformation modeling differs from that of other organ deformation modeling. To be specific, unlike viscoelastic organs such as the liver and brain, spine deformation arises from the inherent flexibility of spinal structures \cite{kadoury2011automatic}. A spine can be segmented into several sections, and the deformation of the entire spine can be dissected into the deformation of each section (as illustrated in Figure \ref{Spine model}). This concept is analogous to ``breaking a curve into multiple straight lines". For each segmented section, traditional rigid registration algorithms, such as bidirectional ICP \cite{du2016robust}, can be employed for modeling the deformation within each specific segment. As the registration progresses in each section, it culminates in local-to-global spine registration. 

To reduce radiation exposure risks for both patients and medical staff, binocular structured light has the potential to acquire intraoperative spinal information instead of relying on traditional X-ray and CT scans. \citet{chen2021method} utilized intraoperative spinal information obtained through binocular structured light to drive the aforementioned multi-section registration method, thereby facilitating spinal deformity correction while minimizing radiation exposure risks during surgery. Nevertheless, it is worth noting that their experiment was conducted in an in-vitro environment. Considering the difference in spine surface exposure extent between in-vitro and real surgical scenarios, further exploration of alternative imaging modalities remains imperative.

\subsection{Vascular surgery}

Intraoperative interventions in vascular surgery encounter intricate vessel anatomy and demand real-time visualization. Currently, these interventions are primarily guided by intraoperative digital substraction angiography (DSA) images \cite{yoon2021topological}. However, the lack of critical 3D information within these images poses challenges for interventional radiologists in accurately identifying vessel branches, catheter tips, and the precise locations of narrowed coronary arteries \cite{zhu2021iterative}. Furthermore, the introduction of rigid endovascular devices exacerbates this challenge by inducing deformations that misalign intended treatments with actual interventions, potentially affecting treatment success and patient outcomes \cite{haigron2013angiovision}.

\begin{figure}[!htb]
\centering
\includegraphics[width=1.0\textwidth]{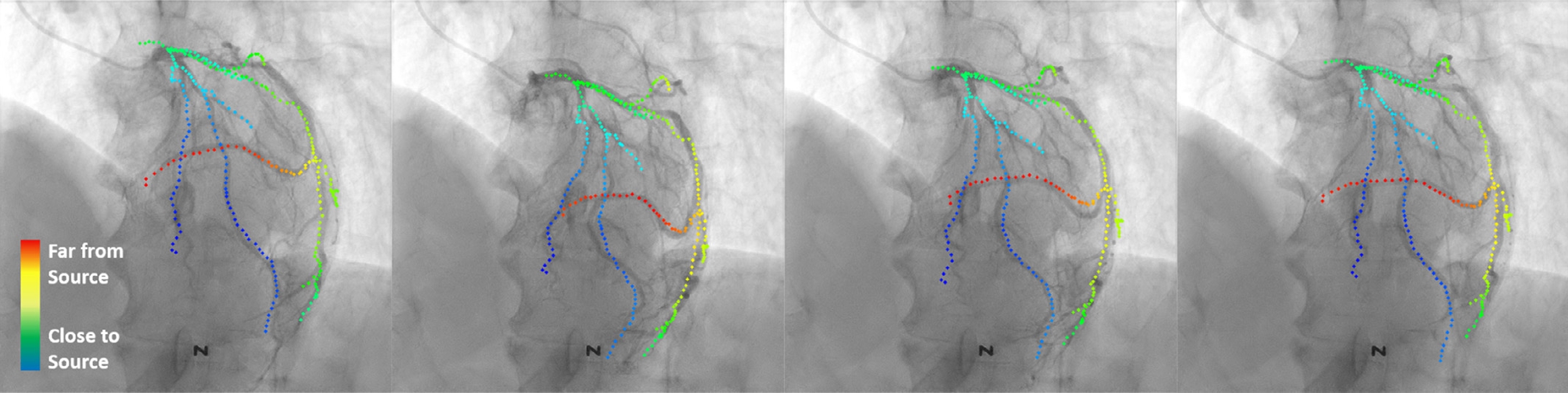}
\caption{Example of vascular surgery with non-rigid registration of 3D vascular models and 2D DSA images \cite{yoon2021topological}. From left to right is the registration of different frames. Permissions: images reprinted with permission from Elsevier.}
\label{Vessel registration}
\end{figure}

To address the challenges mentioned above, deforming preoperative 3D vascular models and fusing them with intraoperative 2D DSA images show great promise \cite{haigron2013angiovision}. This process can be achieved through 2D/3D non-rigid registration methods, where the goal is to iteratively minimize the center-line distance between the 3D vascular structures and the 2D vascular segments visible in DSA images. To tackle the cross-modal registration challenge between 3D vessel models and 2D DSA images, a graph-based representation of vessel tree geometry and topology \cite{el2023intraoperative} can be employed. Subsequently, graph matching algorithms, such as the adaptive compliance graph matching method \cite{garcia2020elastic} or the iterative closest graph matching method \cite{zhu2021iterative}, can be applied to facilitate vascular registration. This approach finds applications in various vascular surgery scenarios, such as percutaneous coronary intervention for abdominal aortic aneurysms, or optimizing drug delivery and tumor targeting during intra-arterial therapies \cite{gerard2017geometric}.

A recent 2D/3D non-rigid registration approach for vascular surgery was introduced by \citet{yoon2021topological}. They achieved a registration accuracy of $1.98 mm$ with respect to the distance error between 3D vessels and the 2D extracted vessels center-line, all while maintaining a computational efficiency of 0.54 seconds. This holds significant promise, as it suggests that surgeons can obtain precise visualizations and navigation within patient-specific vascular structures, ultimately leading to improved treatment strategies and enhanced patient outcomes.

\subsection{Renal surgery}

Modeling organ deformation for surgical navigation is crucial in renal surgery, particularly in the context of minimally invasive procedures \cite{nosrati2015simultaneous}. However, research dedicated specifically to organ deformation modeling in renal surgery remains relatively limited. To some extent, the methods for modeling renal deformation can draw insights from those employed in liver deformation modeling. This is because the challenges in modeling deformation in renal surgery and hepatobiliary surgery exhibit similarities. For instance, both renal and hepatic surfaces lack distinct texture features, and both organs undergo viscoelastic deformations \cite{paulus2017handling}.

A study by \citet{zhang2019markerless} introduced a coarse-to-fine registration framework aimed at applying the coherent point drift algorithm to rectify kidney deformations in laparoscopic partial nephrectomy navigation. Subsequent optimization in their study \cite{zhang2020assessment} improved the accuracy and robustness of this framework against substantial deformations, supported by quantitative evaluations. The average root-mean-square error of volume deformation measured at $0.84 mm$, and the mean navigation TRE from phantom experiments stood at $1.69 mm$.

\subsection{Summary of deformation modeling in surgical specialties}

In summary, the application of deformation modeling methods across various surgical specialties presents a diverse landscape with varying levels of research activity and implementation.

Hepatobiliary surgery takes a prominent position, with a large number of studies (n = 44, p = 41\%) focusing on liver resection and tumor ablation scenarios. While the radiofrequency ablation has achieved noteworthy precision (TRE less than $5mm$) in animal experiments \cite{shi2022synergistic,li2019mixed,li2019augmented}, studies concerning liver resection are still in the in-silico validation stage. Further exploration is needed to establish standardized clinical evaluation protocols and develop modeling algorithms capable of accurately predicting the complex behavior of soft tissues when subjected to large deformation \cite{pfeiffer2019learning}.

In brain surgery (n = 17, p = 16\%), the contours of cortical vessels, exposed on the brain parenchyma surface, offer distinct features that are instrumental in achieving high-precise deformation modeling. Deformation modeling has demonstrated notable accuracy in mitigating disparities resulting from brain shift in-silico settings. Approaches involving non-rigid registration and the incorporation of biomechanical constraints have shown promising levels of accuracy, particularly in cortical and sub-cortical structures, achieving an impressive TRE below $1.93mm$ \cite{haouchine2021pose}. Future endeavors should focus on further improving the accuracy of deeper brain regions and addressing real clinical situations characterized by pronounced degrees of deformation.

Breast surgery (n = 6, p = 5\%), spine surgery (n = 6, p = 5\%), vascular surgery (n = 11, p = 10\%), and renal surgery (n = 4, p = 3\%) have also witnessed the application of deformation modeling methods, albeit with relatively more limited research activity and clinical implementation. In breast surgery, the deformation of breast can be predicted by the displacement of the US probe itself during compression. Data-driven models have shown promise in addressing challenges associated with mammography and ultrasound-guided procedures, showcasing an average TRE of $6.3mm$ in clinical validation \cite{richey2022computational}. In spine surgery, a spine can be segmented into `n' sections, and the deformation of the entire spine can be dissected into the deformation of each section. Multi-stage ICP approaches effectively harness the inherent structural characteristics of the spine to model intraoperative spine deformations, reducing reliance on traditional imaging methods for surgical navigation and thereby decreasing patient radiation exposure \cite{chen2021method}. Vascular surgery utilizes deformation modeling to enhance spatial visualization and navigation within patient-specific vascular structures, leading to improved interventional procedures. This is achieved through the 2D/3D non-rigid registration of preoperative 3D models and intraoperative 2D images, resulting in an average distance error of $1.98 mm$ \cite{yoon2021topological}. In renal surgery,  \citet{zhang2020assessment} have introduced registration frameworks to rectify kidney deformations during laparoscopic procedures, showcasing encouraging prospects for clinical applications, supported by a mean navigation TRE of $1.69 mm $ as evidenced from phantom experiments.

Beyond these, pulmonary surgery (n = 4, p = 3\%), orthopedic surgery (n = 4, p = 3\%), and prostate surgery (n = 4, p = 3\%) emerge with limited studies delving into the application of deformation modeling. While holding promise, these fields are still in an evolutionary phase, warranting further exploration to delineate the most fitting utilization of deformation modeling techniques across their diverse scenarios.

Table \ref{tb:quantitative accuracy results1} and Table \ref{tb:quantitative accuracy results2} provide a summary of deformation modeling studies that have reported quantitative accuracy results. The metrics used to measure the performance of organ deformation modeling algorithms typically include TRE, Hausdorff distance (HD), and root mean square (RMS) error. However, it essential to recognize that the assessment of accuracy is subject to various influencing factors. For instance, \citet{pfeiffer2019learning} have highlighted a direct correlation between algorithm accuracy and the extent of organ deformation. Substantial deformations, such as tissue pulling, often lead to larger absolute errors in algorithm predictions. Conversely, scenarios with minimal organ deformation, such as needle puncture, tend to yield more accurate predictions. Furthermore, the specific organs being studied and the utilization of different intraoperative observations can also directly impact algorithm accuracy. Therefore, it is prudent to consider the reported accuracy measurements as references for algorithm performance rather than the sole criteria for assessment.

\section{\hl{Discussion on Current Status and Future Works}}

Organ deformation modeling serves as a dependable foundation for maintaining the consistency of preoperative organ models with the complex and dynamic intraoperative environment during AR-guided surgery. Overall, the current research field is still at a relatively preliminary stage, with remaining technical challenges to be tackled before possibility of wide adoption in clinical applications.

Existing algorithms for modeling organ deformations can be broadly classified into model-based, data-driven, and hybrid methods. Among these, biomechanical FE-models in model-based methods best explain the organ deformation behavior, and have been successfully applied to simulate the viscoelastic behavior of soft tissues. However, implementing these models directly in real surgical settings presents challenges due to the prolonged computational time required. Specifically, numerical methods for solving the partial differential equations associated with biomechanical FE-models result in solving a system of linear equations. These linear systems are large, sparse, and often ill-conditioned, making traditional numerical solvers inefficient. To address this, leveraging learning-based approaches to obtain high-performing preconditioners offers a promising solution. The preconditioner should transform the original system into an equivalent one with more favorable properties for numerical solving, thereby making the system easier and faster to solve.

We consider exploring the integration of human guidance into the deformation modeling process shows promise for the next step. Currently, deformation modeling methods typically rely on pre-established surface correspondences. These methods estimate the movement of visible surfaces (those with established correspondences) with biomechanical constraints to propagate local motions for inferring the movement of non-visible surfaces. Therefore, establishing accurate correspondences is crucial for precise organ deformation estimation. However, organ surfaces often exhibit highly similar geometric features, such as the left and right lobes of the liver, leading algorithms to converge to incorrect correspondences and affecting the accuracy of subsequent deformation modeling. Existing algorithms struggle to handle such mismatches, especially for textureless surfaces. Incorporating human prior knowledge with carefully designed interaction and user interface in AR has the potential to address this challenge. The key to implementing this approach lies in finding the suitable interactive methods and efficiently leveraging the local reliable correspondence information in practice. Moreover, exploring methods to extend single-frame human annotations into sequential multi-frame data is an intriguing topic. Once humans provide reliable local surface correspondences for a specific deformation state, it is important to continuously utilize such information for subsequent deformations to reduce geometric feature ambiguity.

\hl{Furthermore, we also think that improving the accuracy performance of data-driven algorithms with limited data is a topic that warrants collective discussion in this field.} The limited availability of data stems from the necessity of performing CT scans on patients during surgery to obtain ground-truth deformation of patient organs, thereby increasing the risk of radiation exposure for patients \cite{pfeiffer2020non}. Although there are currently publicly available datasets (cf. Table \ref{tb:Public available datasets}) providing 3D organ models of patients in both preoperative undeformed and intraoperative deformed states, the dataset size is limited. Existing works mainly utilize the valuable data for quantitative evaluation in experiments rather than for training neural networks. There is a strong desire in this field to collect and release more data for research use. Synthetic data with ground truth deformation field are equally appreciated and popular at current status. Data efficient machine learning methods, such as transfer learning, self-supervised learning, foundation models, test-time adaptation have potentials for alleviating the problem. 

Last but not least, ensuring patient safety through rigorous validation and testing protocols is essential. Currently, the accuracy of organ deformation modeling is primarily validated using phantoms or animal studies. Risk factors would be incresingly considered and more frequently discussed along with the advancement of pre-clinical validation of the AR navigation systems. Moreover, bias is a commonly reported issue in data-driven methods, potentially leading to unfair treatment and predicted outcomes. In organ deformation modeling, bias may arise from differences in patient's organ morphology, severity of diseases, surgeon's subjectiveness and incomplete understanding of deformation patterns. The reason behind this is also associated with the afore-mentioned issue of data scarcity. Finally, in relation to the ethical standards of intelligent navigation systems, if that are to be applied, building trust in medical technology is crucial for its adoption. Breaking down the organ deformation prediction process into explainable, intuitive and sequential steps, rather than using a single end-to-end black box, can boost confidence in its clinical applications. This approach could allow surgeons to take over at any unsatisfactory step, leading to more reliable computer-assisted intervention as a whole. Addressing these intertwined issues requires multidisciplinary collaboration and sustained efforts to ensure safety, fairness, and ethical compliance in new technologies for healthcare.

\section{Conclusion}

This literature review presents a systematic and focused overview of methods for modeling organ deformation within the context of AR-guided surgery. We adopt a technology-driven narrative to trace the development and evolution of these modeling techniques. Furthermore, we extend our exploration to encompass various surgical specialties, shedding light on the current state of clinical applications for organ deformation modeling methods while identifying potential barriers to widespread clinical adoption. Building upon existing research, our aim is to provide insights into the future works of organ deformation modeling technology, with an emphasis on unresolved issues that require attention to enhance its applicability. Through this review, we hope to provide readers with an up-to-date understanding of the organ deformation modeling techniques which is fundamental for AR-guided surgery, while also raising interest of future investigations and contributions on this topic.

\section*{Funding}
This project was supported in part by the Science, Technology and Innovation Commission of Shenzhen Municipality (Project No. SGDX20220530111201008), in part by National Natural Science Foundation of China (Project No. 62322318) and in part by the Research Grants Council of Hong Kong Special Administrative Region, China (Projects No. N\_CUHK410/23 and No. T45-401/22-N).

\section*{Disclosure of interest}
No potential conflict of interest need to be reported by the authors in this publication.

\bibliography{interactnlmsample.bib}

\newpage
\section*{Appendix}

\setcounter{table}{0}
\renewcommand{\thetable}{A.\arabic{table}}

\begin{table*}[!htb]
\centering
\caption{The search terms used for the review process.}

\resizebox{\linewidth}{!}{
\begin{tabular}{lp{11cm}l}
\toprule
Sources& Search Terms& Counts\\
\midrule
Scopus& TITLE-ABS-KEY ( ( preoperative  OR  pre-operative  OR  intraoperative  OR  intra-operative  OR  respirat*  OR  breath* )  AND  ( model*  OR  organ  OR  *tissue )  AND  ( *rigid  OR  deform*  OR  motion  OR  displace* )  AND  ( *surg*  OR  laparoscop*  OR  endoscop*  OR  arthroscop*  OR  "Radiofrequency Ablation"  OR  resection  OR  biopsy )  AND  ( navigat*  OR  guid* )  AND  ( regist*  OR  align*  OR  correlat* ) )  AND  ( LIMIT-TO ( LANGUAGE ,  "English" ) )& 1115\\ 

Web of Science& (((((((AB=(Preoperative OR Pre-operative OR Intraoperative OR Intra-operative OR Respirat* OR Breath* OR Interact*)) AND AB=(Mesh OR “Point Cloud” OR Model* OR Organ OR tissue*OR Anatom*)) AND ALL=(nonrigid OR non-rigid OR motion)) AND ALL=(Deform*))) AND AB=(surg* OR Laparoscop* OR Endoscop* OR Arthroscop* OR “Radiofrequency Ablation” OR Resection OR Biopsy)) AND ALL=(Navigat* OR Guid*)) AND ALL=(Regist* OR Correlat*)& 170\\

IEEE Xplore& ("Abstract":Preoperative OR "Abstract":Pre-operative OR "Abstract":Intraoperative OR "Abstract":Intra-operative OR "Abstract":Respirat* OR "Abstract":Breath*) AND ("Abstract":Mesh OR“Point Cloud”OR Model* OR "Abstract":*tissue* OR "Abstract":Organ OR "Abstract":Anatom*) AND ("Full Text Only":*rigid) AND ("Abstract":*surg* OR "Abstract":Laparoscop* OR "Abstract":Endoscop* OR "Abstract":Arthroscop* OR "Abstract":“Radiofrequency Ablation” OR "Abstract":Resection OR "Abstract":Biopsy) AND ("Full Text Only":Navigat* OR "Full Text Only":Guid*) AND ("Full Text Only":Regist* OR "Full Text Only":Align* OR "Full Text Only":Correlat*) AND ("Full Text Only":Deform*)& 265\\

ScienceDirect& Find articles with these terms: (Intraoperative OR Intra-operative) AND (non-rigid OR nonrigid) AND (Deformation OR Motion) AND ( Registration OR Correlation) AND navigation
Title, abstract or author-specified keywords: Surgery OR Laparoscopic OR Endoscopic OR Arthroscopic OR Ablation OR Resection OR Biopsy& 338\\

PubMed& ((((Preoperative[Title/Abstract] OR Pre-operative[Title/Abstract] OR Intraoperative[Title/Abstract] OR Intra-operative[Title/Abstract] OR Respirat*[Title/Abstract] OR Breath*[Title/Abstract]) AND (Model*[Title/Abstract] OR Organ[Title/Abstract] OR *tissue[Title/Abstract])) AND (*rigid OR motion)) AND (deform*)) AND (Navigat* OR Guid* OR Regist* OR Align* OR Correlat*)& 736\\

\hl{Google Scholar} & \hl{(Intraoperative OR Intra-operative) AND (non-rigid OR nonrigid) AND (Deformation OR Motion) AND ( Registration OR Correlation)} & \hl{495}\\

  \bottomrule
  \end{tabular}}
  \label{tbl:table_appendix}
\end{table*}

\setcounter{table}{0}
\renewcommand{\thetable}{\arabic{table}}

\newgeometry{margin=1cm} 
\thispagestyle{empty}

\begin{longtblr}
[
caption        = {Summary of model-based methods for organ deformation modeling. \hl{(The studies with official implementation code are marked with an asterisk (*).)}},
label          = {tb:Model-based deformation modeling methods},
]
{
colspec        = {>{\relsize{-2}}X[0.05,l] >{\relsize{-2}}X[2.0,l] >{\relsize{-2}}X[0.55,l] >{\relsize{-2}}X[1.8,l] >{\relsize{-2}}X[1.5,l]},
hline{1,Z}     = {wd=.08em},
hline{2}       = {wd=.05em},
row{1}         = {font=\bfseries},
rowhead        = 1,
rowsep         = 0.5pt, 
}
& Study & Year & Modeling methods & Organs \\

\SetCell[r=15]{c} \begin{sideways}\textbf{Geometry-based alignment}\end{sideways} & \citet{smit2024ultrasound}, \citet{zhang2020assessment}, \citet{zhang2019markerless}, \citet{allan2016non} & 2024, 2020, 2019, 2016 & Coherent point drift & Liver, kidney, kidney, liver \\
& \citet{zhangnon}, \citet{sun2023research}, \citet{ma2022augmented} & 2024, 2023, 2022 & Non-rigid iterative closest points & Liver, soft tissues, liver \\
& \citet{huang2023non} & 2023 & Hybrid mixture modeling & Liver/lung \\
& \citet{xu2023multi} & 2023 & Multi-constraint point set registration & Coronary artery \\
& \citet{song2023iterative} & 2023 & Iterative Perspective-n-Point &  Femoral artery \\
& \citet{zhu20233d}\hl{*} & 2023 & Manifold regularized modeling  & Coronary artery \\
& \citet{chen2021method} & 2021 & Multi-stage iterative closest points & Spine \\
& \citet{chen2021augmented} & 2021 & Tissue properties-based modeling & Knee\\ 
& \citet{yoon2021topological} & 2021 & Hierarchical registration & Coronary artery \\ 
& \citet{chen2020tissue} & 2020 & Two-level surface warping & Liver\\ 
& \citet{garcia2020elastic} & 2020 & Adaptive compliance graph matching & Vascularized organs \\ 
& \citet{gerard2017geometric} & 2017 & Image-based affine registration & Liver \\
& \citet{maris2017deformable}, \citet{lee2010determining} & 2017, 2010 & Thin-plate spline registration & Breast, lung\\
& \citet{nosrati2015simultaneous} & 2015 & Affine mapping transformation & Kidney\\
& \citet{lange2004augmenting} & 2004 & Multilevel B-splines & Liver\\
\hline
\SetCell[r=21]{c} \begin{sideways}\textbf{Biomechanical models}\end{sideways}& \citet{yang2024boundary} & 2024 & BCs-free FE-model & Liver \\
& \citet{ringel2023regularized} & 2023 & Regularized Kelvinlet model & Breast \\
& \citet{richey2022computational} & 2022 & FE-model & Breast \\
& \citet{golse2021augmented}, \citet{peterlik2018fast}, \citet{adagolodjo2018marker}, \citet{collins2017improving} & 2021, 2018, 2018, 2017 & Co-rotational FE-model & Liver \\ 
& \citet{espinel2020combining} & 2020 & Iterative closest point & Liver \\ 
& \citet{haouchine2020alignment} & 2020 & FE-model & Brain\\ 
& \citet{jobidon2019biomechanically} & 2019 & FE-model & Spine \\ 
& \citet{adagolodjo2017silhouette}, \citet{paulus2017handling} & 2017, 2017 & Co-rotational FE-model & Liver/kidney \\ 
& \citet{plantefeve2016patient}, \citet{haouchine2014impact} & 2016, 2014 & FE-model, beam elements & Liver \\
& \citet{kong2017robust}, \citet{wild2016robust} & 2017, 2016 & FE-model & Kidney \\
& \citet{haouchine2016using}, \citet{wu2014registration}, \citet{haouchine2013deformation}, \citet{rucker2013mechanics} & 2016, 2014, 2013, 2013 & FE-model & Liver \\
& \citet{marques2015framework}, \citet{haouchine2012physics} & 2015, 2012 & Co-rotational FE-model & Liver \\
& \citet{mohammadi2015estimation}, \citet{pheiffer2014model}, \citet{fan2013retractor} & 2015, 2014, 2013 & FE-model & Brain\\ 
& \citet{conley2015realization}, \citet{carter2012application} & 2015, 2012 & FE-model & Breast \\
& \citet{khallaghi2015biomechanically}*, \citet{fedorov2015open}* & 2015, 2015 & Gaussian mixture model \& FE-model & Prostate \\
& \citet{mountney2014augmented} & 2014 & Co-rotational FE-model & Abdominal organs \\
& \citet{berkels2013joint} & 2013 & FE-model & Abdominal and thoracic skin \\
& \citet{haigron2013angiovision} & 2013 & FE-model & Abdominal aortas \\
& \citet{li2013modeling} & 2013 & FE-model & Lung \\
& \citet{zhuang2011sparse}, \citet{liu2007surface}, \citet{bucki2007framework}, \citet{wittek2005brain} & 2011, 2007, 2007, 2005& FE-model & Brain\\ 
& \citet{dumpuri2010model}, \citet{cash2007concepts} & 2010, 2007 & FE-model & Liver \\
\hline
\SetCell[r=7]{c} \begin{sideways}\textbf{Statistical models}\end{sideways}& \citet{guezou2023anterior} & 2023 & Statistical shape model & Pelvis \\
& \citet{zhu2021iterative} & 2021 & Statistical shape model & Vascularized organs  \\
& \citet{li2019mixed,li2019augmented} & 2019 & Statistical motion model & Liver \\ 
& \citet{zhou2018real} & 2018 & Statistical shape model & Liver \\ 
& \citet{hu20153d} & 2015 & Kriging based model morphing& Bone \\
& \citet{kadoury2011automatic} & 2011 & Statistical shape model & Spine \\ 
& \citet{han2009model} & 2009 & Statistical deformation model & Kidney \\
\hline
\SetCell[r=4]{c} \begin{sideways}\textbf{Physics}\end{sideways} & \citet{chen2024method}, \citet{liu2021real} & 2024, 2021 & Position-based dynamics modeling & Liver, soft tissues \\
& \citet{reichard2017projective}, \citet{suwelack2014physics}\hl{*} & 2017, 2014 & Physics-based shape matching & Liver/spleen, liver \\ 
& \citet{costa2012novel} & 2012 & A quasi-static solution & Breast/liver \\
& \citet{dagon2008framework} & 2008 & A mass-spring based modeling & Liver \\

\end{longtblr}
\restoregeometry 

\newgeometry{margin=1cm} 
\thispagestyle{empty}
\begin{longtblr}
[
caption        = {Summary of data-driven and hybrid methods for organ deformation modeling. \hl{(The studies with official implementation code are marked with an asterisk (*).)}},
label          = {tb:Data-driven and hybrid methods},
]
{
colspec        = {>{\relsize{-2}}X[0.05,l] >{\relsize{-2}}X[1.5,l] >{\relsize{-2}}X[0.5,l] >{\relsize{-2}}X[2.6,l] >{\relsize{-2}}X[1.1,l]},
hline{1,Z}     = {wd=.08em},
hline{2}       = {wd=.05em},
row{1}         = {font=\bfseries},
rowhead        = 1,
rowsep         = 0.5pt, 
}
& Study & Year & Modeling methods & Organs \\

\SetCell[r=19]{c} \begin{sideways}\textbf{Hybrid algorithms}\end{sideways} & 
\SetCell[c=4]{l}\textbf{NetBiom} & & & \\
& \citet{wang2024libr+} & 2024 & Spline-residual GCN $+$ Biomechanical model & Liver \\
& \citet{labrunie2023automatic} & 2023 & CNN $+$ 3D morphable models & Liver \\
& \citet{mendizabal2023intraoperative}\hl{*} & 2023 & BANet2.0 $+$ Biomechanical model & Liver \\ 
& \citet{tagliabue2021intra}\hl{*} & 2021 & BANet $+$ Biomechanical model & Liver \\

& \SetCell[c=4]{l}\textbf{MultiDef} & & & \\
& \citet{ha20242d} & 2024 &  CNN $+$ Statistical shape model & Femur \\
& \citet{liebmann2024automatic} & 2024 & U-Net $+$ Multi-stage iterative closest points & Spine \\
& \citet{zhang2023spr} & 2023 & Vision Transformer $+$ Thin-plate spline & Coronary artery \\
& \citet{shi2022synergistic} & 2022 & U-Net $+$ Generalized moving least-square algorithm & Liver \\
& \citet{haouchine2021pose,haouchine2020deformation} & 2021, 2020 & U-Net $+$ Force-based shape-from-template formulation & Brain \\ 
& \citet{jia2021improving} & 2021 & Point cloud occupancy network $+$ Nonlinear optimization & Liver \\ 
& \citet{lei2021diffeomorphic} & 2021 & $\epsilon$-support vector regression $+$ Kriging algorithm & Liver/lung \\ 
& \citet{shakeri2021deformable} & 2021 & Deep variational autoencoder $+$ Non-rigid optimal ICP & Prostate\\
& \citet{shi2021internal} & 2021 & U-Net $+$ Moving least-squares algorithm & Liver \\

& \SetCell[c=4]{l}\textbf{PhysDL} & & & \\
& \citet{odot2023real} & 2023 & MLP $+$ Optimal control & Liver \\
& \citet{min2023non}\hl{*} & 2023 & Physics-informed DL & Prostate \\
& \citet{salehi2022physgnn}\hl{*} & 2022 & PhysGNN & Brain \\
\hline
\SetCell[r=17]{c} \begin{sideways}\textbf{Data-driven algorithms}\end{sideways} & 
\SetCell[c=4]{l}\textbf{Machine learning approaches} & & & \\
& \citet{el2024towards} & 2024 & U-Net & Liver \\ 
& \citet{azampour2023anatomy}\hl{*} & 2023 & FlowNet3D & Spine \\ 
& \citet{young2023investigating} & 2023 & U-Net $+$ Vision Transformer & Liver \\ 
& \citet{shao2023real} & 2023 & GCN $+$ U-Net & Liver \\ 
& \citet{nakao2022image}\hl{*} & 2022 & Image-to-graph convolutional network & Abdominal organ \\ 
& \citet{yamamoto2021kernel} & 2021 &  Kernel regression & Lung \\ 
& \citet{pfeiffer2020non}\hl{*} & 2020 & U-Net & Liver \\ 
& \citet{pfeiffer2019learning}\hl{*} & 2019 &  U-Net & Liver \\ 
& \citet{mendizabal2020physics} & 2019 & U-Net & Breast \\
& \citet{brunet2019physics} & 2019 & U-Net & Liver \\
& \SetCell[c=4]{l}\textbf{Altas-based modeling} & & & \\
& \citet{heiselman2018characterization} & 2018 & Altas-based modeling & Liver \\
& \citet{miga2016clinical} & 2016 & Altas-based modeling & Brain \\
& \citet{chen2013integrating} & 2013 & Altas-based modeling & Brain \\
& \citet{chen2010intraoperative} & 2010 & Altas-based modeling & Brain \\
& \citet{clements2007atlas} & 2007 & Altas-based modeling & Liver\\

\end{longtblr}
\restoregeometry 

\newgeometry{margin=1cm} 
\thispagestyle{empty}
\begin{longtblr}
[
caption        = {\hl{Comparison of different deformation modeling methods}},
label          = {tb:Comparison of different deformation modeling methods},
]
{
colspec        = {>{\relsize{-2}}X[1.8,c] >{\relsize{-2}}X[1,c] >{\relsize{-2}}X[1,c] >{\relsize{-2}}X[1,c] >{\relsize{-2}}X[1,c] >{\relsize{-2}}X[1,c] >{\relsize{-2}}X[1,c] >{\relsize{-2}}X[1,c] >{\relsize{-2}}X[1,c] >{\relsize{-2}}X[1,c] >{\relsize{-2}}X[1,c]},
hline{1,Z}     = {wd=.08em},
hline{2}       = {wd=.05em},
row{1}         = {font=\bfseries},
vline{2,6,9} = {1,2,3,4,5}{0.05em},
vline{3,4,5,7,8,10,11} = {2}{0.01em},
rowhead        = 1,
rowsep         = 0.5pt, 
}
\SetCell[r=2]{c} Criterions & \SetCell[c=4]{c} Model-based methods & & & & \SetCell[c=3]{c} Data-driven methods & & & \SetCell[c=3]{c} Hybrid methods & & \\
& Biom. & Phys. & Geom. & Stat. & Atlas & TML & DL & NetBiom & MultiDef & PhysDL \\
\hline
Accuracy & High & Medium & Low & High & Low & Medium & Low & Medium & Medium & Medium \\
Efficiency & Low & Medium & High & High & High & High & High & Medium & Low & High \\
Implementation & Low & High & High & Low & High & Low & High & High & High & High \\

\end{longtblr}

\begin{longtblr}
[
caption        = {\hl{Public available datasets for organ deformation modeling}},
label          = {tb:Public available datasets},
]
{
colspec        = {>{\relsize{-2}}X[1.9,l] >{\relsize{-2}}X[0.9,l] >{\relsize{-2}}X[0.9,l] >{\relsize{-2}}X[0.9,l] >{\relsize{-2}}X[0.8,l] >{\relsize{-2}}X[1.1,l] >{\relsize{-2}}X[0.7,l]},
hline{1,Z}     = {wd=.08em},
hline{2}       = {wd=.05em},
row{1}         = {font=\bfseries},
rowhead        = 1,
rowsep         = 0.5pt, 
}
Datasets & Preoperative organ models & Deformed organ models & Intraoperative organ images & Validation targets & Organ & Subjects \\

Sparse Data Challenge \cite{heiselman2024image} & \checkmark & \checkmark &  & \checkmark & Phantom liver & 1\\

Open-CAS LiverReg \cite{suwelack2014physics} & \checkmark & \checkmark &  & \checkmark & Phantom liver & 1\\

CALLR \cite{rabbani2022methodology} & \checkmark & & \checkmark & \checkmark & Liver & 4 \\

DePoLL \cite{modrzejewski2019vivo} & \checkmark & \checkmark & \checkmark & \checkmark & Porcine liver & 1 \\

Respiratory Motion \cite{De_Jong2019-sj} & \checkmark & \checkmark &  &  & Liver & 4 \\

3D-IRCADb-02 \cite{3dircadb02} & \checkmark & \checkmark &  &  & Abdominal organ & 1 \\

P2ILF (coming soon) \cite{ali2024objective} & \checkmark &  & \checkmark &  & Liver & 11 \\

C3VD \cite{bobrow2023} & \checkmark &  & \checkmark &  & Colon & 1 \\

Med-RAD \cite{fried2023dataset} & \checkmark & \checkmark &  &  & Lung & 3 \\

Vascular Tree \cite{xu2023multi} & \checkmark & \checkmark &  &  & Coronary artery & 16 \\

Prostate-MRI-US-Biopsy \cite{natarajan2020prostate} & \checkmark & \checkmark &  & \checkmark & Prostate & 642 \\

Tissue Dissection \cite{tagliabue2021intra} & \checkmark & \checkmark &  & & Silicone phantoms & 4 \\

\end{longtblr}

\restoregeometry 

\newgeometry{margin=1cm} 
\thispagestyle{empty}
\begin{longtblr}
[
caption        = {Quantitative accuracy results reported in organ deformation modeling studies. \hl{(The studies with official implementation code are marked with an asterisk (*).)}},
label          = {tb:quantitative accuracy results1}
]
{
colspec        = {>{\relsize{-2}}X[0.5,l] >{\relsize{-2}}X[2.7,l] >{\relsize{-2}}X[0.8,l] >{\relsize{-2}}X[2.3,l] >{\relsize{-2}}X[1.8,l] >{\relsize{-2}}X[2,l] >{\relsize{-2}}X[2.7,j]},
hline{1,Z}     = {wd=.08em},
hline{2}       = {wd=.05em},
row{1}         = {font=\bfseries},
rowhead        = 1,
}
& Study  & Year & \hl{Imaging modality} & Experiments & Metrics & Results (unit: mm) \\

\SetCell[r=31]{c} \begin{sideways}Hepatobiliary Surgery\end{sideways} & \citet{yang2024boundary} & 2024 & RGB-D images & Phantom & Average TRE & 3.14 $\pm$ 1.34\\
& \citet{wang2024libr+} & 2024 & RGB-D images & In-silico & Average TRE & 3.23 $\pm$ 0.76 \\
& \citet{chen2024method} & 2024 & Structured light & Ex-vivo & TRE & 1.24 \\
& \citet{smit2024ultrasound} & 2024 & US images & Clinical cases & Median TRE & 6.90 \\
& \citet{mendizabal2023intraoperative}\hl{*} & 2023 & RGB-D images &In-vivo & RMS TRE & 0.87\\
& \citet{young2023investigating} & 2023 & Stereo laparoscopy &In-silico & Average TRE & 4.70 $\pm$ 0.50\\
& \citet{shao2023real} & 2023 & X-ray images & In-silico & 95-percentile HD & 2.40 $\pm$ 1.60 \\
& \citet{labrunie2023automatic} & 2023 & Monocular laparoscopy & In-silico & Average TRE & 8.50 $\pm$ 3.90 \\
& \citet{huang2023non} & 2023 & RGB-D & Phantom & Average TRE & 8.58 $\pm$ 0.84 \\
& \citet{ma2022augmented} & 2022 & US images & 1). Phantom;\newline 2). Ex-vivo & Average TRE & 1). 2.34 $\pm$ 0.45;\newline 2). 4.49 $\pm$ 1.88. \\
& \citet{shi2022synergistic} & 2022 & Optical trackers & In-vivo & TRE & 2.50 \\
& \citet{golse2021augmented} & 2021 & RGB-D images & Ex-vivo & RMS TRE & 7.90 \\
& \citet{jia2021improving} & 2021 & RGB-D images & In-silico & Average TRE & 4.28 $\pm$ 0.60 \\
& \citet{espinel2020combining} & 2020 & Monocular laparoscopy & Phantom & Average TRE & 5.53 $\pm$ 1.41 \\
& \citet{chen2020tissue} & 2020 & US images & Ex-vivo & Average TRE & 2.14 $\pm$ 0.21 \\
& \citet{pfeiffer2020non}\hl{*} & 2020 & Stereo laparoscopy & In-silico & TRE & TRE$_{mean}$ 5.70 \\
& \citet{li2019mixed} & 2019 & Optical trackers & In-vivo & Average TRE & 3.52 $\pm$ 0.12 \\
& \citet{li2019augmented} & 2019 & Optical trackers & In-vivo & Average TRE & 2.68 $\pm$ 1.02\\
& \citet{brunet2019physics} & 2019 & RGB-D images & Ex-vivo & TRE & TRE$_{mean}$ 2.92 \\ 
& \citet{peterlik2018fast} & 2018 & CBCT images & In-silico & TRE & TRE$_{mean}$ 3.80 \\
& \citet{heiselman2018characterization} & 2018 & Spatial resampling & Phantom & Average TRE & 6.70 $\pm$ 1.30. \\
& \citet{collins2017improving} & 2017 & Spatial resampling & Phantom & Average TRE & 5.30 $\pm$ 0.50 \\
& \citet{allan2016non} & 2016 & Stereo laparoscopy & Phantom & RMS TRE & 2.89 \\
& \citet{haouchine2014impact} & 2014 & Stereo endoscopy & Phantom & TRE & 4.41 \\
& \citet{haouchine2014towards} & 2014 & Stereo endoscopy & Ex-vivo & TRE & 5.13\\
& \citet{suwelack2014physics}\hl{*} & 2014 & RGB-D images & Phantom & TRE & 8.70 \\ 
& \citet{plantefeve2014automatic} & 2014 & Stereo laparoscopy & In-vivo & Mean HD & 0.60 \\ 
& \citet{haouchine2013image} & 2013 & Stereo endoscopy & Phantom & TRE & 4.40 \\ 
& \citet{dumpuri2010model} & 2010 & RGB-D images & Clinical cases & Average TRE & 3.10 $\pm$ 1.80 \\ 
& \citet{cash2007concepts} & 2007 & RGB-D images & Phantom & RMS TRE & 6.70\\
& \citet{cash2005compensating} & 2005 & Laser range scans & Phantom & RMS TRE & 2.40\\
\hline
\SetCell[r=9]{c} \begin{sideways}Brain Surgery\end{sideways} 
& \citet{salehi2022physgnn}\hl{*} & 2022 & RGB-D images & In-silico & TRE & 0.31 $\pm$ 0.69\\
& \citet{haouchine2021pose} & 2021 & 2D images & In-silico & TRE & TRE$_{max}$ 1.93 \\
& \citet{haouchine2020alignment} & 2020 & 2D images & In-silico & TRE & TRE$_{max}$ 2.00 \\
& \citet{miga2016clinical} & 2016 & RGB-D images & Clinical cases & TRE &  TRE$_{max}$ 4.00 \\
& \citet{pheiffer2014model} & 2014 & Laser range scans & Clinical cases &  Average TRE & 2.00 $\pm$ 0.90 \\
& \citet{zhuang2011sparse} & 2011 & Laser range scans & In-vivo & Average TRE & 1.62 $\pm$ 0.22 \\
& \citet{liu2010novel} & 2010 & Laser range scans & In-vivo & TRE &  1.20 \\
& \citet{liu2007surface} & 2007 & Stereo-vision images & In-vivo & Surface distance &  1.38 \\
& \citet{wittek2005brain} & 2005 & MRIs & In-silico & TRE & 3.18. \\
\end{longtblr}
\restoregeometry 

\newgeometry{ margin=1cm}
\thispagestyle{empty}
\begin{longtblr}
[
caption        = {Quantitative accuracy results reported in organ deformation modeling studies. \hl{(The studies with official implementation code are marked with an asterisk (*).)}},
label          = {tb:quantitative accuracy results2}
]
{
colspec        = {>{\relsize{-2}}X[0.5,l] >{\relsize{-2}}X[3.0,l] >{\relsize{-2}}X[0.8,l] >{\relsize{-2}}X[2.3,l] >{\relsize{-2}}X[1.8,l] >{\relsize{-2}}X[2.3,l] >{\relsize{-2}}X[2.4,j]},
hline{1,Z}     = {wd=.08em},
hline{2}       = {wd=.05em},
row{1}         = {font=\bfseries},
rowhead        = 1,
}
& Study  & Year & \hl{Imaging modality} & Experiments & Metrics & Results (unit: mm) \\

\SetCell[r=9]{c} \begin{sideways}Vascular Surgery\end{sideways} 
& \citet{zhang2023spr} & 2023 & DSA images & Clinical cases & Average HD & 1.73 $\pm$ 0.62 \\
& \citet{song2023iterative} & 2023 & DSA images & Clinical cases & Projected distance & 3.74 \\
& \citet{zhu20233d}\hl{*} & 2023 & DSA images & In-silico & Projected distance & 0.34 \\
& \citet{yoon2021topological} & 2021 & DSA images & In-silico & Projected distance & 1.98 \\
& \citet{zhu2021iterative} & 2021 & DSA images & Clinical cases & Projected distance & 1.59 $\pm$ 0.40 \\
& \citet{garcia2020elastic} & 2020 & CBCT images & In-silico & TRE & 4.20 \\
& \citet{garcia2018biomechanics} & 2018 & CBCT images & In-silico & Average TRE & 4.00 $\pm$ 2.10 \\
& \citet{gerard2017geometric} & 2017 & US images & In-silico & Average TRE & 3.90 $\pm$ 1.10 \\
& \citet{haigron2013angiovision} & 2013 & DSA images & Clinical cases & Average TRE & 2.90 $\pm$ 0.50 \\
\hline
\SetCell[r=6]{c} \begin{sideways}Spine Surgery\end{sideways} & \citet{liebmann2024automatic} & 2024 & RGB-D images & Ex-vivo & Average TRE & 1.20 $\pm$ 0.22 \\
& \citet{azampour2023anatomy}\hl{*} & 2023 & US images & In-silico & Average TRE & 3.67 $\pm$ 0.63 \\
& \citet{chen2021method} & 2021 & Structured light & Ex-vivo & RMS TRE & 0.51 $\pm$ 0.31  \\
& \citet{jobidon2019biomechanically} & 2019 & CBCT images & Clinical cases & 1). Surface distance;\newline 2). Orientation error. & 1). 1.50 $\pm$ 1.20;\newline 2). 2.7 $\pm$ 2.6\degree.\\
& \citet{kadoury2011automatic} & 2011 & CT images & Clinical cases & RMS TRE & 1.80 $\pm$ 0.70 \\
& \citet{rajamani2007statistical} & 2007 & US images & In-silico & Mean HD & 1.76 \\
\hline
\SetCell[r=4]{c} \begin{sideways}Pulmonary Surgery\end{sideways} & \citet{huang2023non} & 2023 & RGB-D images & Clinical cases & Average TRE & 5.30 $\pm$ 0.37 \\
& \citet{yamamoto2021kernel} & 2021 & Sampled landmarks & In-silico & 1). RMS TRE;\newline 2). HD. & 1). 2.74;\newline 2). 6.11. \\
& \citet{li2013modeling} & 2013 & CT images & Clinical cases & Average TRE & 1.30 $\pm$ 0.97 \\
& \citet{lee2010determining} & 2010 & Laser range scans & Phantom & Average TRE & 2.32 $\pm$ 0.23 \\
\hline
\SetCell[r=12]{c} \begin{sideways}\makecell{Breast Surgery, Orthopedic Surgery, \\  Renal Surgery, Prostate Surgery}\end{sideways} 
& \citet{ringel2023regularized} & 2023 & RGB-D images & Clinical cases & Average TRE & 3.00 $\pm$ 1.10 \\
& \citet{richey2022computational} & 2022 & MRIs & In-vivo & Average TRE & 4.20 $\pm$ 1.00 \\
& \citet{mendizabal2020physics} & 2020 & Optical trackers  & Phantom & Average TRE & 4.21 $\pm$ 1.01 \\
& \citet{ha20242d} & 2024 & X-ray images & Clinical cases & Average HD & 4.43 $\pm$ 0.85 \\
& \citet{guezou2023anterior} & 2023 & US images & In-vivo & Orientation error & 7.3 $\pm$ 3.5\degree \\
& \citet{chen2021augmented} & 2021 & Sampled landmarks & 1). Phantom;\newline 2). Ex-vivo. & Average TRE & 1). 2.01 $\pm$ 0.65;\newline 2). 2.97 $\pm$ 0.79. \\
& \citet{hu20153d} & 2015 & Spatial resampling & Phantom & RMS TRE & 0.75. \\
& \citet{zhang2020assessment} & 2020 & Stereo endoscopy & Phantom & Average TRE & 1.69 $\pm$ 0.31 \\
& \citet{zhang2019markerless} & 2019 & Stereo endoscopy & Phantom & RMS TRE & 1.28 \\
& \citet{min2023non}\hl{*} & 2023 & TRUS images & Clinical cases & Average TRE & 6.12 $\pm$ 1.95 \\
& \citet{shakeri2021deformable} & 2021 & TRUS images & In-silico & Average TRE & 3.90 $\pm$ 1.40 \\
& \citet{khallaghi2015biomechanically}* & 2015 & TRUS images & Clinical cases & Average TRE & 2.89 $\pm$ 1.44 \\
\hline
\SetCell[r=6]{c} \begin{sideways}Other Surgical Procedures\end{sideways} & \citet{nakao2022image}\hl{*} & 2022 & X-ray images & In-silico & 1). Surface distance;\newline 2). HD. & 1). 2.12 $\pm$ 0.84;\newline 2). 10.27 $\pm$ 4.02.\\
& \citet{lei2021diffeomorphic} & 2021 & CT images & In-vivo & TRE & 2.10 \\
& \citet{tagliabue2021intra}\hl{*} & 2021 & RGB-D images & Ex-vivo & RMS TRE &  3.62 \\
& \citet{pfeiffer2019learning}\hl{*} & 2019 & RGB-D images & Phantom & TRE & TRE$_{mean}$ 5.10\\
& \citet{adagolodjo2017silhouette} & 2017 & 2D images & Phantom & TRE & 2.20 \\
& \citet{paulus2017handling} & 2017 & 2D images & Phantom & Maximum HD & 6.90\\
\end{longtblr}
\restoregeometry

\end{document}